\documentclass[reprint, amsmath,amssymb, aps, prl, superscriptaddress]{revtex4-2}

\usepackage{graphicx}% Include figure files
\usepackage{dcolumn}% Align table columns on decimal point
\usepackage{bm}
\usepackage{multirow}
\usepackage{amsfonts}
\usepackage{amssymb}
\usepackage{amsthm}
\usepackage{array}
\usepackage{amsmath}
\usepackage{verbatim}
\usepackage{hyperref}
\usepackage{color}
\usepackage{bbold}
\usepackage{epstopdf}
\usepackage{mathtools}
\usepackage{MnSymbol}
\usepackage{tikz}
\usepackage[export]{adjustbox}
\usepackage{subfigure}
\usepackage{threeparttable}

\newcommand{\affa}{State Key Laboratory of Low Dimensional Quantum Physics, Department of Physics, Tsinghua University, Beijing 100084, China}
\newcommand{\affb}{Beijing Academy of Quantum Information Sciences, Beijing 100193, China}
\newcommand{\affc}{Hefei National Laboratory, Hefei 230088, P. R. China}
\newcommand{\affd}{Frontier Science Center for Quantum Information, Beijing 100084, China}

\newcommand{\Yb}{$^{171}\rm{Yb}^+$}

\newcommand{\Sr}{$^{88}\rm{Sr}^+$}

\newcommand{\ddd}[1]{{\color{blue}#1}}

\newcommand{\ket}[1]{\left\vert#1\right\rangle}
\newcommand{\bra}[1]{\left\langle#1\right\vert}

\def\bra#1{\langle #1|}
\def\ket#1{\left|#1 \right>}

\begin{document}

\title{Entangling gates for trapped-ion quantum computation and quantum simulation}

\author{Zhengyang Cai}
\address{\affa}
\author{Chunyang-Luan}
\address{\affa}
\author{Lingfeng Ou}
\address{\affa}
\author{Hengchao Tu}
\address{\affa}
\author{Zihan Yin}
\address{\affa}
\author{Jing-Ning Zhang}
\address{\affb}
\author{Kihwan Kim}
\email{kimkihwan@mail.tsinghua.edu.cn}
\address{\affa}
\address{\affb}
\address{\affc}
\address{\affd}

\begin{abstract}
The trapped-ion system has been a leading platform for practical quantum computation and quantum simulation since the first scheme of a quantum gate was proposed by Cirac and Zoller in 1995. Quantum gates with trapped ions have shown the highest fidelity among all physical platforms. Recently, sophisticated schemes of quantum gates such as amplitude, phase, frequency modulation, or multi-frequency application, have been developed to make the gates fast, robust to many types of imperfections, and applicable to multiple qubits. Here, we review the basic principle and recent development of quantum gates with trapped ions. 
\end{abstract}

\maketitle

\section{I. Introduction} 
In 1995, the first quantum gate for quantum computation was proposed and realized with trapped ions \cite{cirac1995quantum,monroe1995demonstration,schmidt2003realization}. Since the proposal by Cirac and Zoller~\cite{cirac1995quantum}, the trapped ion system has led the field of quantum computation~\cite{leibfried2003quantum,haffner2008quantum,ladd2010quantum,blatt2012quantum,monroe2013scaling} as well as quantum simulation~\cite{monroe2021programmable}. The Cirac-Zoller (CZ) gate is based on specifically using the center of mass vibrational mode. The CZ gate requires perfect ground-state cooling and individual addressing on the target ions, which makes it difficult to be a high-fidelity gate. In 1999, Mølmer and Sørenson proposed a different type of entangling gate, which relaxes the requirements of perfect ground state cooling and individual addressing, and the gates were realized with up to four qubits by NIST group \cite{sorensen1999quantum,molmer1999multiparticle,sorensen2000entanglement,sackett2000experimental}. Almost at the same time, a gate using geometric phase in the vibrational mode space from conditional displacement operation was proposed and realized with decent fidelity ~\cite{solano1999deterministic,milburn2000ion,leibfried2003experimental}, which was also named as light-shift (LS) gate. It has a similar level of experimental requirements to the Mølmer-Sørenson (MS) gate. Later, the MS gate was understood in terms of the geometric phase in the phase space of position and momentum, which unifies the understanding of the MS gate and LS gate \cite{sorensen2000entanglement,lee2005phase}.

Basically, both the LS \cite{leibfried2003experimental} and the MS gates can be understood as the specific usage of a force dependent on a qubit state. For the LS gate \cite{leibfried2003experimental}, the force is dependent on the ion-qubit state in the $\sigma_{\rm z}$ basis, and for the MS gate, it is dependent on the ion-qubit state in the $\sigma_{\phi}$-basis, where $\sigma_{\phi}= \cos \phi~\sigma_{\rm x}+ \sin \phi ~\sigma_{\rm y}$ \cite{sorensen2000entanglement,lee2005phase}. Therefore, the LS gate and the MS gate are called $\sigma_{\rm z}$- and $\sigma_{\phi}$-gate, respectively. The LS and the MS gate have been improved and shown the fidelities of over 99.9 $\%$, which have been the highest fidelities so far among all physical platforms for quantum computation \cite{gaebler2016high,ballance2016high,clark2021high}.

One of the promising schemes to scale up the size of the ion-trap system is to entangle a small number of ion-qubits by using the LS or the MS gate in a single zone and to connect different trapping zones by individually shuttling ions \cite{kielpinski2002architecture,metodi2005quantum}, which is called a quantum charge-coupled device (QCCD) architecture. These entangling gates are typically realized by single or Raman laser beams with non-vanishing wave-vector $\mathbf{ k}$. The laser implementations of the trapped-ion quantum gates have been successful with a small number of ion-qubits with high fidelities \ddd{\cite{gaebler2016high,ballance2016high,clark2021high}}. However, it could be challenging to apply numerous laser beams to different trapping zones for large-scale quantum computation based on the QCCD approach. The microwave or RF implementation could provide an alternative solution for implementing quantum gates on multiple zones since microwave or RF circuits can be integrated with the trap~\cite{ospelkaus2008trapped,ospelkaus2011microwave}. These laser-less gates require a significant field gradient to couple qubits and vibrational modes. Two main methods have been explored in this direction; one is using static-magnetic-field gradient combined with microwave~\cite{mintert2001ion,timoney2011quantum,khromova2012designer,weidt2016trapped}, and the other one is using oscillating-magnetic-field gradient at near-qubit frequency~\cite{ospelkaus2008trapped,ospelkaus2011microwave,harty2016high}. Recently, the third scheme has been implemented, which combines a near-motion-frequency field gradient and near-qubit-frequency field~\cite{ding2014microwave,srinivas2019trapped,srinivas2021high}.  

Alternatively, it has been proposed and realized to perform entangling gates on any two qubits in a large number of ions in a single trap zone, where the total number of vibrational modes increases with the number of ions in the zone~\cite{zhu2006trapped,zhu2006arbitrary,lin2009large,choi2014optimal,debnath2016demonstration}. In this case, it is necessary to take into account the effect of many collective vibrational modes, which is to disentangle target qubits and all relevant vibrational modes at the end of the gate. Originally, these requirements were proposed to be fulfilled by using amplitude modulation~\cite{zhu2006trapped,zhu2006arbitrary,lin2009large,steane2014pulsed}. Later, other methods such as phase~\cite{green2015phase,milne2020phase,bentley2020numeric} or frequency modulation~\cite{leung2018robust,leung2018entangling,landsman2019two}, and multi-tone modulation~\cite{haddadfarshi2016high,webb2018resilient,shapira2018robust,shapira2020theory}, have been introduced for the purpose and others such as robustness or noise-resilience of the pairwise gates~\cite{haddadfarshi2016high}. Moreover, these sophisticated methods of quantum gates can be also applied to make two-qubit gates without the speed limit~\cite{steane2014pulsed}. Furthermore, these schemes were extended to simultaneously or parallelly entangle more qubits~\cite{lu2019global,figgatt2019parallel,shapira2020theory,wang2022fast}. 

In this article, we provide the basic principle of quantum entangling gates with trapped ions and review the recent development of gates with sophisticated pulse modulations for various purposes. We point out that among many different types of modulations, the multi-frequency method captures the general control that covers all other modulation methods and provides systematic ways of optimizing the gates according to the purposes. Finally, we summarize some applications of the multi-frequency method for various purposes. %speed-up, robustness, and simultaneous operation on multiple ions. 
This review article consists of the following sections. In section II, we introduce the basics of the trapped-ion system, such as different types of qubits, cooling, qubit initialization, detection, and single-qubit operations. In section III, we discuss the interactions between a two-level system coupled to a single vibrational mode, entangling gate operations with and without laser beams based on the qubit-state dependent force. In section IV, we summarize the modulation methods such as amplitude, phase, frequency, and multi-frequency, and show that the multi-frequency method has the most general aspect of control. In section V, we provide a general theoretical framework and examples of the multi-frequency method with or without individual controls to speed up the gate, make the gate robust against various noises and apply the gate on more than two ions simultaneously. Finally, we conclude the review article with an outlook.

\section{II. Basics of The Trapped-Ion System}
\subsection{Types of ion qubits}

\begin{figure}
\centering
\includegraphics[width=0.95\linewidth]{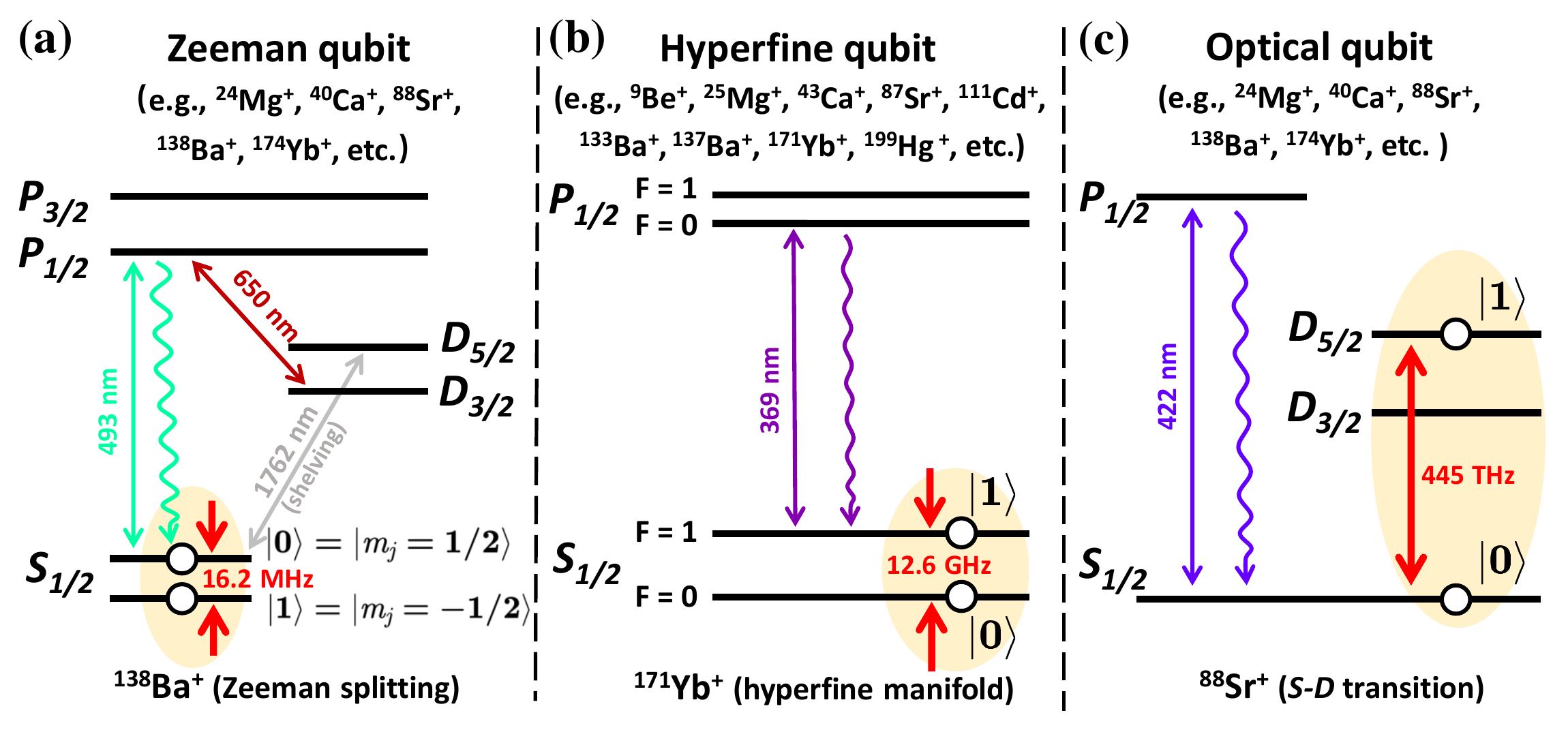}
%\caption{The scheme of three typical types of ion-qubits. (a) Zeeman qubit, for example $^{138}\rm{Ba}^+$ ions. (b) Hyperfine qubits, for example $^{171}\rm{Yb}^+$ ions. (c) Optical qubits, for example $^{88}\rm{Sr}^+$ ions.}
\caption{The level structure of three typical types of ion-qubits (energy splittings not to scale). (a) The level splitting of Zeeman qubit is on the order of a few to tens of megahertz (as in $^{138}\rm{Ba}^+$, etc.). (b) Hyperfine qubit, composed of a pair of ground-state hyperfine levels, has the level splitting on the order of a few to tens of gigahertz (as in $^{171}\rm{Yb}^+$, etc.). (c) Consisting of one ground state and another metastable level, the optical qubit is driven by a light field on the order of a few hundred terahertz (as in $^{88}\rm{Sr}^+$, etc.).}
\label{fig:qubit size}
\end{figure}

Served as a qubit, two internal levels of a trapped atomic ion are typically encoded as $\ket{0}$ and $\ket{1}$. Different from other artificial qubits, ion qubits are fundamentally identical, which is an advantageous feature for large-scale system development. The ion qubits have more unique advantages over other qubits in different physical platforms, such as the ultra-long coherence time up to the order of hours~\cite{langer2005long,harty2014high,wang2017single,wang2021single} and near-perfect qubit-state initialization and detection~\cite{myerson2008high,todaro2021state,harty2014high}. There are typically three types of qubits, where the energy splitting covers the frequency range of a few to tens of megahertz, gigahertz, and hundreds of terahertz, which correspond to the Zeeman qubit (Fig.~\ref{fig:qubit size}(a)) \cite{ruster2016long}, the hyperfine qubit (Fig.~\ref{fig:qubit size}(b))~\cite{leibfried2003experimental,blinov2004quantum,benhelm2008towards,olmschenk2007manipulation,ballance2016high} and the optical qubit (Fig.~\ref{fig:qubit size}(c)), respectively.

%Zeeman qubit is composed of a pair of electronic states with the same orbital and hyperfine energy level as shown in the Fig.~\ref{fig:qubit size}(a), and it can be manipulated by the radio-frequency(RF) field\cite{keselman2011high,chu2021precise} or Raman lasers\cite{poschinger2009coherent,ruster2016long}. Due to the simpler energy level structure of Zeeman qubit, the state initialization, Doppler cooling and sideband cooling can be performed directly. However, the state discrimination of Zeeman qubit requires another metastable energy level to perform the shelving operation\cite{blatt2004ion,dietrich2010hyperfine}.

The Zeeman qubit is composed of a pair of 
Zeeman energy levels as shown in Fig.~\ref{fig:qubit size}(a), and they can be manipulated by applying the radio-frequency(RF) or Raman lasers with frequency difference in the order of MHz~\cite{ruster2016long}. For the Zeeman qubit, Doppler cooling, sideband cooling, and state initialization can be simply implemented without any advanced experimental requirements such as high-frequency sidebands, narrow-line lasers, etc. However, the high-efficiency state discrimination of the Zeeman qubit requires another metastable energy level to shelve one of the qubit states. Moreover, the Zeeman qubit is sensitive to magnetic-field fluctuations, which results in a relatively short coherence time on the order of milliseconds. With the magnetic field shielding and permanent magnets, the coherence time of the Zeeman qubit has been increased to $\sim 2.1$ s with dynamical decoupling pulses~\cite{ruster2016long}.
For the optical qubit, one level in a ground state manifold and the other level in a metastable electronic state are used, which are typically separated by the electric quadrupole or octupole transition, as shown in the Fig.~\ref{fig:qubit size}(c). The transitions can be driven with a single-frequency laser in the visible to near-IR spectrum region. Due to the long lifetime of metastable levels, the narrow-line laser ($\sim 1$ Hz) stabilized by the high-finesse cavity is required to perform the optical transition. The coherence time of the optical qubit is similar to that of the Zeeman qubit.  %relatively short ($\sim 0.2$ s)~\cite{bermudez2017assessing}, 
Technical efforts have been attempted to suppress phase fluctuations of the laser and push the coherence time to the ultimate limit of the upper-level decay~\cite{bermudez2017assessing}.

%For optical qubits, the energy splitting between the ground-state manifold and the metastable level is typical a few hundreds of terahertz, and the electric quadrupole transition can be performed with a narrow-line laser beam in the visible to near-infrared spectrum region\cite{poschinger2009coherent,clark2021high,wang2022significant}. However, the high-finesse cavity with a line-width of a few hertz is required to stabilize the narrow-line laser for the optical transition. Due to the larger energy splitting, the state detection of optical qubits has higher efficiencies and the $S_{1/2}-D_{5/2}$ manifolds transition is typical used as the shelving operation as discussed above. 

%Thus, the suppression of magnetic field fluctuations can be realized by the magnetic field shielding and permanent magnets with low temperature drift\cite{ruster2016long, wang2021single}, and the coherence time can be increased to a few hundreds of millisecond\cite{ruster2016long}. In addition, the relationship between the external magnetic field and the decoherence rate of Zeeman qubits can be served as the magnetic field quantum sensor\cite{wei2022measurement}.

Consisting of a pair of ground-state hyperfine levels, the hyperfine qubit can be driven by the microwave field~\cite{shappert2013spatially,harty2016high} or Raman lasers with the frequency difference in the order of GHz, as shown in Fig.~\ref{fig:qubit size}(b). The hyperfine qubit can be implemented in “clock" transitions that are insensitive to the first-order magnetic field and have shown coherence time up to the order of hours~\cite{langer2005long,harty2014high,wang2017single,wang2021single}. 
Moreover, the dressed-states composed of ground-state hyperfine levels coupled with resonant RF field~\cite{timoney2011quantum,randall2015efficient} is an available qubit to realize a laser-less gate with a magnetic field gradient.

\subsection{Cooling, state initialization, and detection}
To obtain high-fidelity quantum entangling gates with trapped ions, motional-ground state cooling is necessary, which is realized by laser cooling methods. First, based on the velocity-dependent radiation force~\cite{wineland1978radiation,neuhauser1978optical}, Doppler cooling can typically cool down ions to the order of millikelvin, which is limited by the natural linewidth of the cooled ions. Near ground-state cooling can be achieved by resolved-sideband cooling~\cite{monroe1995resolved,roos1999quantum}. Moreover, Sisyphus cooling~\cite{dalibard1989laser,ejtemaee20173d}, or electromagnetically-induced-transparency (EIT) cooling provides an alternative to cooling a large number of modes of multiple ions simultaneously~\cite{morigi2000ground,roos2000experimental,lin2013sympathetic,lechner2016electromagnetically, scharnhorst2018experimental,jordan2019near,feng2020efficient,qiao2021double}.

Before applying sequences of quantum gate operations, the ion qubits should be initialized to one of the qubit states. The state initialization is typically performed by the optical pumping technique within less than a few microseconds~\cite{wineland1980double}. %To date, the highest fidelity (99.93 $\%$) of state preparation has demonstrated with the $^{43} \rm{Ca}^+$ ions~\cite{harty2014high} \ddd{(check more references).} 
%To date, the highest 99.999 $\%$ fidelity of state initialization has demonstrated with $^{138} \rm{Ba}^+$ ion~\cite{christensen2020high}.
At the end of the quantum operations, the states of ion qubits can be measured via the state-dependent fluorescence detection~\cite{bergquist1986observation,nagourney1986shelved}. The ion qubit is projected to a bright state $\ket{1}$, which scatters lots of photons with the illumination of a detection laser, or a dark state $\ket{0}$ that scatters almost no photons. The scattered fluorescence photons are collected by an imaging system and detected by a photomultiplier tube (PMT) or charge-coupled device (CCD) typically. Recently, over 99.99 $\%$ detection fidelities have been demonstrated with optical qubits of $^{40} \rm{Ca}^+$ and $^{138} \rm{Ba}^+$~\cite{myerson2008high,zhukas2021high}. For the hyperfine qubit of a single $^{133} \rm{Ba}^+$ ion, around 99.97 $\%$ detection fidelity has been achieved~\cite{christensen2020high}. For other hyperfine qubits with $^{43} \rm{Ca}^+$ and $^{171} \rm{Yb}^+$, over 99 $\%$ detection fidelities have been observed~\cite{myerson2008high,crain2019high}. The duration for the state detection typically takes around a few hundred $\mu$s to milliseconds. The shortest detection time was reported to $\sim$ 11 $\mu$s with 99.93 $\%$ average fidelity achieved by superconducting nanowire single-photon detectors (SNSPDs)~\cite{crain2019high}.

%with less than 200 $\mu$s detection time~\cite{myerson2008high}, and the high-speed detection ($\sim 10 ~\mu$s) with 99.3$\%$ fidelity is also achieved~\cite{crain2019high}. \ddd{include more references} Recently, up to 99.971(3) $\%$ detection fidelity has been achieved in a single $^{133} \rm{Ba}^+$ ion by PMT methods~\cite{christensen2020high}. And a single $^{138} \rm{Ba}^+$ ion detection using CCD methods with over 99.999 $\%$ fidelity was also been demonstrated~\cite{zhukas2021high}. In addition, the high-speed detection ($\sim$ 11 $us$) with 99.931(6) $\%$ average fidelity was also achieved by superconducting nanowire single-photon detectors (SNSPDs)~\cite{crain2019high}. 

\subsection{Single-qubit operations}

%For a single ion-qubit, the internal electronic structure can be approximately simplified to a two-level system as shown in the Fig.~\ref{fig: Rabi oscillation}, and the transition between $\ket{0}$ and $\ket{1}$ states is performed by the single-qubit rotation operations\cite{nielsen2002quantum,leibfried2003quantum}. The flopping oscillations between the two-level system with Rabi frequency $\Omega$ can be driven by a near-resonant field, including the resonant RF field\cite{harty2014high,tan2015multi}, Raman lasers\cite{ballance2016high,campbell2010ultrafast} and the single laser\cite{bermudez2017assessing}. The single qubit state is described as $\ket{\psi} = \cos{(\theta/2)} \ket{0} +e^{i \phi} \sin{(\theta/2)} \ket{1}$, where the Rabi frequency is $\Omega = \theta/t$,  and it represents the coupling strength of field-matter interaction\cite{hong2017experimental,matjelo2021demonstration}.

%For a single ion-qubit, the internal electronic structure can be approximately simplified to a two-level system as shown in the Fig.~\ref{fig: Rabi oscillation}, and the transition between $\ket{0}$ and $\ket{1}$ states is performed by the single-qubit rotation operations\cite{nielsen2002quantum,leibfried2003quantum}. 

The single qubit operation can be driven by a resonant field to the qubit frequency, which can be realized by RF field~\cite{harty2014high}, Raman lasers~\cite{ballance2016high,campbell2010ultrafast} or the single laser~\cite{bermudez2017assessing}. The single qubit state is described as $\ket{\psi} = \cos{(\theta/2)} \ket{0} +e^{i \phi} \sin{(\theta/2)} \ket{1}$, where the Rabi frequency is $\Omega = \theta/t$,  and it represents the coupling strength of field-matter interaction~\cite{leibfried2003quantum}.

In the experiment, the highest fidelity of 99.9999 $\%$ single-qubit gates for the hyperfine qubit of $^{43} \rm{Ca}^+$ has been performed by the near-field microwave~\cite{harty2014high}. %\ddd{LCY: I have checked the recent refs: over 99.995 $\%$ fidelity for Phys.Rev.Lett.117,060505(2016)~\cite{gaebler2016high}, and about 99.9934(3) $\%$ for Phys.Rev.Lett.117,060504(2016)~\cite{ballance2016high} }. 
With the Raman laser beams, the fidelities of over 99.99$\%$ have been demonstrated for the hyperfine qubits of $^{43} \rm{Ca}^+$ and $^{9} \rm{Be}^+$~\cite{ballance2016high,gaebler2016high}. For optical qubits, over 99.99$\%$ fidelities of a single qubit gate have been achieved with a single narrow-line laser~\cite{bermudez2017assessing}.

% In order to speed up the single-qubit gate, some technical efforts have been attempted but at the expense of reduced fidelity. For the Raman method, the gate duration has been pushed to less than 50 ps with 99$\%$ fidelity\cite{campbell2010ultrafast}. And even the gate timescales of 20 $ns$ performed by the microwave method is also achieved but no clear fidelity values\cite{ospelkaus2011microwave}. In addition, the fidelity of ultra-fast single-qubit gate can be improved with the application of modulated pulse sequences and the suppression of unwanted excited transitions.

In order to speed up the single-qubit gates, mainly technical efforts have been attempted. For the Raman method, the gate duration has been pushed to less than 50 $ps$~\cite{campbell2010ultrafast,guo2022picosecond}. By using the microwave method, the gate with timescales of 20 $ns$ has been performed~\cite{ospelkaus2011microwave}. In addition, the fidelity of ultra-fast single-qubit gate can be improved with the application of modulated pulse sequences and the suppression of unwanted excited transitions~\cite{mizrahi2013ultrafast,wong2017demonstration}.

For the single qubit operations with ion qubits, in particular, the hyperfine qubits, the main error sources are technical imperfections such as fluctuations in the amplitude of microwave fields. For the optical operations, the photon scattering from the relevant dipole transitions is an additional limiting factor~\cite{ozeri2007errors}, which makes lower fidelities than those of microwave gates. The coupling to vibrational modes in the single-qubit operations can be significantly suppressed by using fields with large wavelengths or copropagating-Raman schemes that nullify the coupling to vibrational modes. The coherence time of clock state qubits already reached over hours~\cite{langer2005long,harty2014high,wang2017single,wang2021single}, which is not already a limiting factor for the fidelity of single-qubit operations.

\section{III. A brief explanation of laser and laserless gates} 

\subsection{Two-level system coupled to a single vibrational mode}

For multi-qubit gates, quantum information stored in the internal states between trapped ions is transferred via coupling to the collective vibrational mode, the so-called “quantum bus", which plays a core role in the quantum entangling  gates~\cite{leibfried2003quantum,lee2005phase,haffner2008quantum}. With the near-resonant field, we just take into account a two-level system and one vibrational mode, and the general interaction Hamiltonian can be expressed as, 
\begin{equation}
\hat{H}_{\rm int}=\frac{\hbar}{2} \Omega [e^{i(k x-\omega t+\phi)}+e^{-i(k x-\omega t+\phi)}] (\hat{\sigma}^{+}+\hat{\sigma}^{-}), 
  \label{eq:interaction coupling Hamiltonian}
\end{equation}
where $\mathbf{k}\equiv k\hat{\mathbf e}_x$, $\omega$ and $\phi$ are the wave vector, frequency, and phase of the field, respectively, $\Omega= \frac{|\mathbf{d} \cdot \mathbf{E}|} {\hbar}$ is the Rabi frequency, and $\hat{\sigma}^+ = \ket{1} \bra{0}$, $\hat{\sigma}^- = \ket{0} \bra{1}$. We assume the field is coupled to the vibrational mode in the $x$-radial direction.
Transformed to the interaction picture with respect to $\hat{H_0} = \frac{\hbar}{2} \omega_{0} \hat{\sigma^z}+ \hbar\omega_{\rm m} \hat{a}^\dag \hat{a}$, the above interaction Hamiltonian (\ref{eq:interaction coupling Hamiltonian}) can be simplified after the rotating wave approximation (RWA),
\begin{equation}
    \hat{H}_{\rm int}^{'}=\frac{\hbar}{2} \Omega \left\{\hat{\sigma}^+ e^{i [\eta (\hat{a}^\dagger e^{i \omega_{\rm m} t} + \hat{a} e^{-i \omega_{\rm m} t}-\mu t +\phi)]} \right\} + \text{h.c.},
  \label{eq:rotating wave approximation}
\end{equation}
where $\mu = \omega-\omega_0$ is the detuning between the field frequency and the qubit frequency $\omega_0$, $\omega_{\rm m}$ is the motional mode frequency, and $\eta = k \sqrt{\frac{\hbar}{2 m \omega_{\rm m}}}$ is the Lamb-Dicke parameter.

\begin{figure}
\centering
\includegraphics[width=0.95\linewidth]{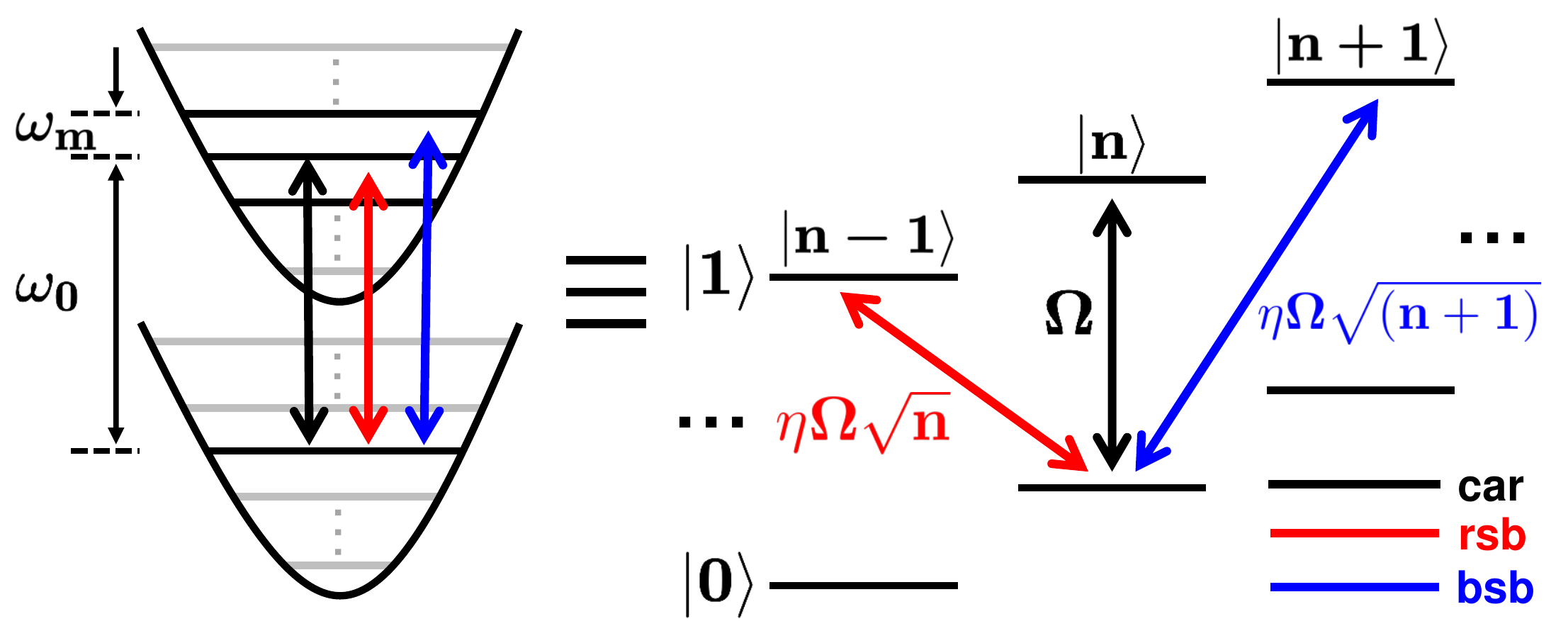}
\caption{The illustration for carrier and sideband transitions, where the $\omega_0$ and $\omega_{\rm m}$ are the qubit frequency and motional mode frequency. The abbreviations of car (black line), rsb (red line) and bsb (blue line) represent carrier, red- and blue-sideband transitions, respectively.}
\label{fig:Carrier and sideband transitions}
\end{figure}

Generally, the phonon number $\ket{n}$ of trapped ions is cooled down to the Lamb-Dicke regime \cite{leibfried2003quantum,lee2005phase,haffner2008quantum}, where $\eta \sqrt{n+1} \ll 1$. And in this regime, the interaction Hamiltonian (\ref{eq:rotating wave approximation}) can be further simplified to the following form,
\begin{equation}
    \hat{H}_{\rm LD}=\frac{\hbar}{2} \Omega \hat{\sigma}^+ \left[1 + i \eta \left(\hat{a}^\dagger e^{i \omega_{\rm m} t} + \hat{a} e^{-i \omega_{\rm m} t} \right) \right] e^{i (\phi - \mu t)} + \text{h.c.}.
  \label{eq:Lamb-Dick regime}
\end{equation}
In particular, we consider the three types of interaction, the carrier transition ($\mu= 0$), detuned blue ($\mu=\omega_{\rm m}+\delta$), and red sideband transitions ($\mu=-\omega_{\rm m}-\delta$),
\begin{align}
    \hat{H}_{\rm car} &= \frac{\hbar}{2} \Omega (\hat{\sigma}^+ e^{i \phi_{\rm c}}+\hat{\sigma}^- e^{-i \phi_{\rm c}}), \nonumber\\
    \hat{H}_{\rm bsb} &= i \eta \frac{\hbar}{2} \Omega (\hat{\sigma}^+ \hat{a}^\dagger e^{i \phi_{\rm b}} e^{-i \delta t} - \hat{\sigma}^- \hat{a} e^{-i \phi_{\rm b}} e^{i \delta t}),\\
    \hat{H}_{\rm rsb} &= i \eta \frac{\hbar}{2} \Omega (\hat{\sigma}^+ \hat{a} e^{i \phi_{\rm r}} e^{i \delta t} - \hat{\sigma}^- \hat{a}^\dagger e^{-i \phi_{\rm r}}e^{-i \delta t}), \nonumber
  \label{eq:car rsb bsb  Hamiltonian}
\end{align}
where $\delta \ll \omega_{\rm m}$. 
As shown in Fig.~\ref{fig:Carrier and sideband transitions}, the carrier transition $\ket{0, n} \leftrightarrow \ket{1,n}$ is driven with Rabi frequency $\Omega$, while the blue ($\ket{0, n} \leftrightarrow \ket{1,n+1}$) and red sideband transitions ($\ket{0, n} \leftrightarrow \ket{1,n-1}$) are driven with the Rabi frequencies of $\eta \Omega \sqrt{n+1}$ and $\eta \Omega \sqrt{n}$, respectively.

For an ion-qubit coupled with a single vibrational mode by a near-resonant field, the qubit-state-dependent force can be achieved by combining the blue and red sideband transitions, 
\begin{eqnarray}
    \hat{H}_{\rm SDF}&&= \hat{H}_{\rm rsb}+\hat{H}_{\rm bsb} \nonumber\\
    &&=\eta \frac{\hbar}{2} \Omega  (\hat{a}^\dagger e^{i \phi_{\rm m}} e^{-i \delta t}+ \hat{a} e^{-i \phi_{\rm m}} e^{i \delta t})\hat{\sigma}^{\phi_{\rm S}},
  \label{eq:spin-dependent force}
\end{eqnarray}
in which we have $\phi_{\rm S} = (\phi_{\rm b}+\phi_{\rm r} - \pi )/2$, $\phi_{\rm m} = (\phi_{\rm b} - \phi_{\rm r})/2$, and $\hat{\sigma}^{\phi_{\rm S}}=\hat{\sigma}^x\cos\phi_{\rm S} + \hat{\sigma}^y\sin\phi_{\rm S}$. It is clear that $\phi_{\rm S}$ and $\phi_{\rm m}$ correlate to the phase of qubit state and vibrational mode, respectively~\cite{haljan2005spin,lee2005phase}. 

For the sake of ubiquity, we can generalize the above Hamiltonian in the following way~\cite{haljan2005spin,lee2005phase},
\begin{equation}
    \hat{H}_{\rm SDF}=i\hbar(\gamma(t)\hat{a}^{\dagger}-\gamma^*(t)\hat{a})\hat{\sigma}^{\phi_{\rm S}},
\end{equation}
and the unitary evolution can be written as 
\begin{equation}
    \hat{U}(t)=\hat{D}(\alpha(t)\hat{\sigma}^{\phi_{\rm S}})\exp(i\Phi(t)),
\end{equation}
where $\hat{D}$ is the displacement operator, and $\Phi(t)$ is the geometric phase~\cite{lee2005phase}. The expressions of $\gamma(t)$, $\alpha(t)$, and $\Phi(t)$ are expressed as the following way, 
\begin{align}
   \gamma(t)&=-i \frac{\Omega}{2}\eta e^{-i \delta t} e^{i \phi_{\rm m}},\nonumber\\
   \alpha(t)&=\int_{0}^{t} \left(-i \frac{\Omega}{2}\eta e^{-i \delta t}e^{i\phi_{\rm m}}\right) dt = \frac{\Omega \eta e^{i\phi_{\rm m}}}{2 \delta}\left(e^{-i \delta t} - 1\right),\\
   \Phi(t)&= {\rm Im} \left(\int_0^t\,\alpha(t')^*\mathrm{d}\alpha(t')\right).
   \nonumber
  \label{eq:parameters}
\end{align}

When $\delta=0$, the spin-dependent force can be simplified to the following expression,
\begin{equation}
   \hat{H}_{\rm SDF}^{\delta=0}=\eta\frac{\hbar}{2} \Omega\left(\hat{a}^\dagger e^{i\phi_{\rm m}} + \hat{a}e^{-i\phi_{\rm m}} \right)\hat{\sigma}^{\phi_{\rm S}},
\end{equation}
where the corresponding terms are $\gamma(t)=-i \Omega \eta e^{i\phi_{\rm m}}/2$, $\alpha(t)=-i \Omega \eta e^{i\phi_{\rm m}} t/2$, which results in no geometric phase as $\Phi(t)=0$.

\begin{figure}
\centering
\includegraphics[width=0.75\linewidth]{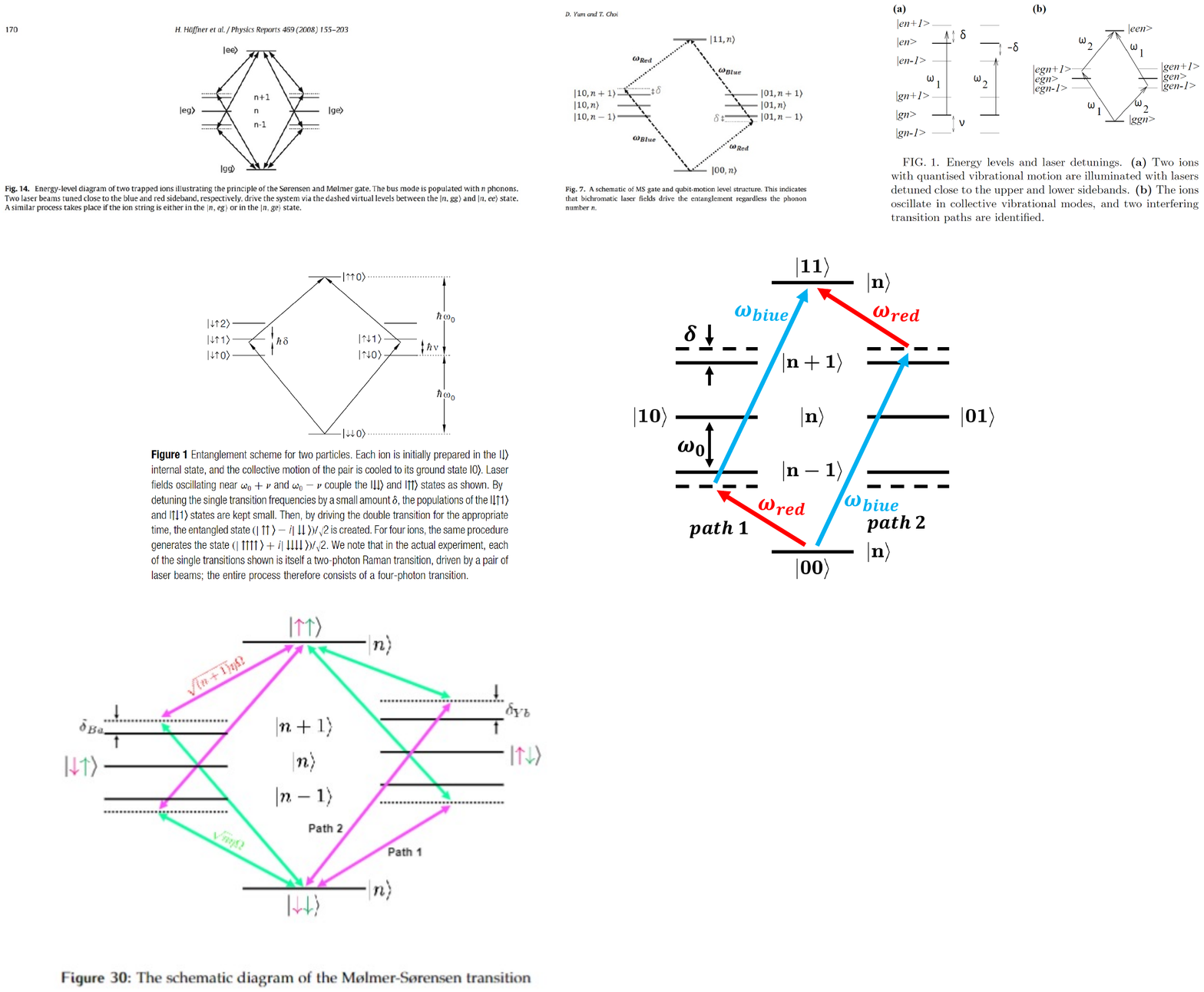}
\caption{The trajectory of motion is driven by the spin-dependent force in the phase space. The state $\ket{+}$ and $\ket{-}$ represent the two eigenstates of $\hat{\sigma}^{\phi_{\rm S}}$ with eigenvalues +1 and -1, respectively.}
\label{fig:phase space}
\end{figure}

However, when the detuning $\delta \neq 0$, the evolution of $\alpha$ will go through a circular trajectory, as is drawn in Fig.~\ref{fig:phase space}. The left and right branches of this figure which circle in different directions corresponding to the positive and negative eigenstate of $\hat{\sigma}^{\phi_{\rm S}}$, will generate the same geometric phases equal to the size of the encapsulated area. When the trajectories loop back to the origin, the motional states are disentangled from the spin states. 

By far, we have illustrated a means of generating $\sigma^{\phi_{\rm S}}$-spin-dependent force, which directly leads to the MS gates. We note that the spin-dependent force can be realized in $\sigma^{\rm z}$-basis as discussed in \cite{milburn2000ion,leibfried2003experimental,lee2005phase}, which leads to the LS gates. 

%various ways, such as the LS gates \cite{milburn2000ion,leibfried2003experimental,lee2005phase, etc...}, which cannot be implemented in field-insensitive qubits in general. It was proposed to realize the LS gates with field-insensitive qubits by using the laser detunings at the middle of trap frequencies~\cite{roos2008ion,kim2008geometric} or narrow line transitions~\cite{aolita2007high}. Recently, these LS gates with clock states have been realized~\cite{monz,honeywell,bazavan2022synthesizing}...   University of Oxford\cite{bazavan2022synthesizing}, etc. All these schemes have formed the fundamental blocks of generating 2-qubit and multi-qubit gates.

\subsection{Laser gate}
In ion trap systems, the two-qubit entangling gates can still be implemented via the coupling between the internal states and the collective vibrational modes. In this section, we mainly focus on using a single collective mode and later we will discuss the general situation with more-than one collective vibrational mode.  

The trapped-ion entangling gates can be realized in various different ways with or without laser beams. Throughout the years, employing combinations of tunable lasers to implement multi-qubit gates has been the most prevalent and adhibited scenario, either for the hyperfine qubits or the optical qubits. The hyperfine qubits are typically incorporated with Raman transitions and the optical qubits are operated by applying direct transitions with narrow-linewidth. 

The original entangling gate between two ion-qubits was proposed by Cirac and Zoller~\cite{cirac1995quantum} and realized with a single qubit and two qubits~\cite{monroe1995demonstration,schmidt2003realization}. However, the CZ gate requires cooling to the ground state of the collective vibrational mode and the laser addressing on individual ions, which made it difficult to realize high-fidelity gates. 

Different from the CZ gate scheme, the Mølmer-Sørenson(MS) gate \cite{sorensen1999quantum,molmer1999multiparticle,sorensen2000entanglement} is independent with phonon number $\ket{n}$ of the vibrational mode of the interest, which is more experimentally favorable than the CZ gate scheme. For the MS gate scheme, a pair of trapped ions are illuminated with bichromatic laser fields simultaneously, and the ions will be driven by the state-dependent force to follow different motional trajectories depending on qubit states in the position-momentum phase space~\cite{sorensen2000entanglement,lee2005phase}. In order to be an entangling operation, the internal states should be disentangled at the end of the gate with the accumulated geometric phase of $\pi /4$. The MS gate scheme was firstly demonstrated with $^{9}\rm{Be}^+$ hyperfine qubits~\cite{sackett2000experimental} in 2000. The MS gates have been popularly demonstrated in many ion-trap groups with up to the fidelity of 99.9~$\%$~\cite{home2009complete,gaebler2016high}. The effective gate-error rates of the two-qubit gates were also suppressed to $(0.96 \pm 0.10) \times 10^{-3}$ from $10^{-2}$ level of physical infidelity by using the error mitigation technique~\cite{zhang2020error}. 

\begin{figure}
\centering
\includegraphics[width=0.75\linewidth]{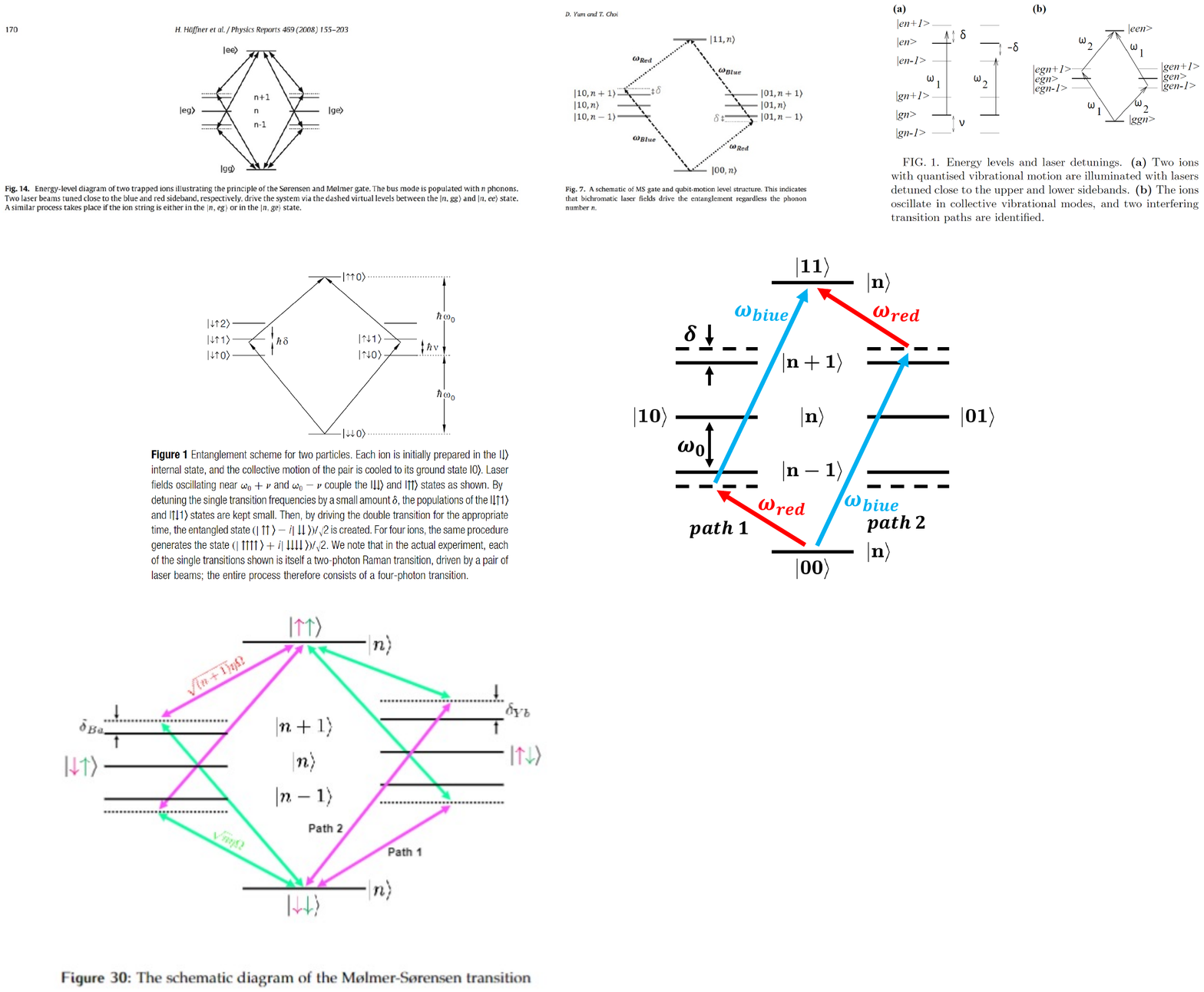}
\caption{The illustration of two typical paths for MS gate, which are driven by the bichromatic laser fields regardless of the phonon number $\ket{n}$, and the $\omega_0$ and $\delta$ are the qubit frequency and laser detuning.}
\label{fig:MS gate}
\end{figure}

For MS gate, the spin-flip process is actualized by using a bichromatic laser whose central frequency corresponds to the single-flip resonance frequency, and the sideband frequency is slightly detuned from the vibrational mode, as Fig.~\ref{fig:MS gate} shows \cite{roos2008ion}. To simplify our discussion, only one axial mode will be included below. In this way, the Hamiltonian may be derived from directly extending Eq.~(\ref{eq:spin-dependent force}) to a two-ion scenario~\cite{lee2005phase},
\begin{equation}
    \hat{H}_{\rm SDF} = \sum_{j=1}^{2}\eta_{j}\frac{\hbar\Omega}{2}(\hat{a}^{\dagger}\mathrm{e}^{-i\delta t}\mathrm{e}^{i\phi_{\mathrm{m},j}}
    +\hat{a}\mathrm{e}^{i\delta t}\mathrm{e}^{-i\phi_{\mathrm{m},j}})\hat{\sigma}^{\phi_{\mathrm{S},j}}_j,
    \label{MS Hamiltonian}
  \end{equation}
where $\eta_j$ and $\phi_{\mathrm{m},j}$ represent the Lamb-Dicke parameter and the phase correlated to the vibrational mode of each ion, respectively. Suppose we have $\phi_{\mathrm{S},1}=\phi_{\mathrm{S},2}=\phi_{\rm S}$, then, the above Hamiltonian will lead to the following evolution operator,
    \begin{gather}
        \hat{U}(t) = \exp\left[\sum_{j=1}^2\left(\alpha_{j}(t)a^{\dagger}-\alpha_{j}^*(t)a\right)\hat{\sigma}^{\phi_{\rm S}}_j
        -i\frac{\theta(t)}{2}\hat{\sigma}^{\phi_{\rm S}}_1\hat{\sigma}^{\phi_{\rm S}}_2\right], \label{eq:yzh_1}
    \end{gather}
where
\begin{align}
    \alpha_{j}(t)&=-i\frac{\eta_j\Omega}{2}\int_0^t\,\mathrm{e}^{-i\delta t}\mathrm{e}^{i\phi_{\mathrm{m},j}}\mathrm{d}t=\frac{\Omega\eta_j\mathrm{e}^{i\phi_{\mathrm{m},j}}}{2\delta}\left(\mathrm{e}^{-i\delta t} - 1\right), \label{eq:alpha}
    \\
    \theta(t)&=\eta_1\eta_2\Omega^2\left(\frac{t}{\delta}-\frac{\sin(\delta t)}{\delta^2}\right).
    \label{eq:theta}
\end{align}

Viewing from phase space, the internal ion states will be fully disentangled with the vibrational mode after the phase trajectories $\alpha_j(t)$ of Eq.~(\ref{eq:alpha}) are closed, and the first term of Eq. (\ref{eq:yzh_1}) disappears. At this time, $\theta(t)$ the geometric phase of the two-ion gate shown in Eq.~(\ref{eq:theta}) is obtained~\cite{sorensen2000entanglement}, and the evolution of MS interaction can be simplified to \cite{manovitz2022trapped},
\begin{equation}
    \hat{U}(\theta)=\exp\left(-i\frac{\theta}{2}\hat{\sigma}_1^{\phi_{\rm S}}\hat{\sigma}_2^{\phi_{\rm S}}\right). 
\end{equation}

Another typical two-qubit entangling gate scheme was proposed by Milburn in 2000~\cite{milburn2000ion} and demonstrated with $^{9}\rm{Be}^+$ hyperfine qubits~\cite{leibfried2003experimental}, which is so-called LS gate. Similar to the MS gate scheme, the LS gate is also driven by the state-dependent force and generates the entangling interaction between ion qubits with the collective vibrational mode. However, the $\ket{0} \leftrightarrow \ket{1}$ states transition is not required during LS gate operation. LS gate is mainly achieved by applying two beams with a frequency difference close to a target vibration mode, which is coupled in the direction along the wavevector difference. In the Lamb-Dicke regime and the rotating wave approximation, the interaction hamiltonian can be written as:
\begin{equation}
    \hat{H}_{\rm I} = \sum_{j=1,2} \sum_{s=0,1} \frac{\hbar\eta_j}{2}
    \Omega_s \ket{s}\bra{s}_j \left[
    \hat{a} e^{i(\delta t-\phi_{\mathrm{m},j})} + \hat{a}^\dagger e^{-i(\delta t-\phi_{\mathrm{m},j})}
    \right].
\end{equation}
When we have $\Omega_0=-\Omega_1=\Omega$, we may transform the Hamiltonian into the following form:
\begin{equation}
    \hat{H}_{\rm SDF} = \sum_{j=1}^{2}\eta_{j}\frac{\hbar\Omega}{2}(\hat{a}^{\dagger}\mathrm{e}^{-i\delta t}\mathrm{e}^{i\phi_{\mathrm{m},j}}
    +\hat{a}\mathrm{e}^{i\delta t}\mathrm{e}^{-i\phi_{\mathrm{m},j}})\hat{\sigma}^{z}_j. 
    \label{eq:LSHam}
\end{equation}

We can immediately discover that the above Hamiltonian of Eq. (\ref{eq:LSHam}) is almost the same as the MS gate Hamiltonian of Eq. ~(\ref{MS Hamiltonian}), except $\hat{\sigma}_{j}^{\phi_{\rm S}}$ is replaced by $\hat{\sigma}_{j}^{z}$, which indicates the similarity of the MS gate and the LS gate. In this case, we may acquire the evolution operator in a form similar to Eq. (\ref{eq:yzh_1}), 
    \begin{gather}
        \hat{U}(t) = \exp\left[\sum_{j=1}^2\left(\alpha_{j}(t)a^{\dagger}-\alpha_{j}^*(t)a\right)\hat{\sigma}_{z}^j
        -i\frac{\theta}{2}(t)\hat{\sigma}^{z}_1\hat{\sigma}^{z}_2\right] \label{eq:yzh_2},
    \end{gather}
where the definition of $\alpha_j(t)$ and $\theta(t)$ directly follow Eqs. (\ref{eq:alpha}, \ref{eq:theta}). When the single qubit interaction terms vanish, we have,
\begin{equation}
    \hat{U}(\theta)=\exp\left(-i\frac{\theta}{2}\hat{\sigma}_1^{z}\hat{\sigma}_2^{z}\right). 
\end{equation}

The imperfection of the LS gate comes from multiple sources, including higher-order terms beyond the Lamb-Dicke regime, laser-control errors, motional decoherence, unwanted mode coupling, scattering error, etc. This type of gate has been demonstrated on hyperfine qubits~\cite{leibfried2003experimental,ballance2016high} and optical qubits~ \cite{sawyer2021wavelength,clark2021high}. Fast gate with fidelity of $99.8\%$ in 1.6~$\mathrm{\mu s}$ has been achieved on $^{43}\mathrm{Ca}^+$ hyperfine qubits~\cite{schafer2018fast}. Bell-State with infidelity of $6(3)\times 10^{-4}$ without subtraction of experimental errors has been directly measured with $^{40}\text{Ca}^+$ optical qubits~\cite{clark2021highfidelity}. 

The LS gate was considered to be impossible to implement with field-insensitive qubits~\cite{lee2005phase}. However, it was proposed to realize the LS gates with field-insensitive qubits by using the laser detunings at the middle of trap frequencies~\cite{roos2008ion,kim2008geometric} or narrow line transitions~\cite{aolita2007high}. These LS gates with clock states have been realized~\cite{monz2009realization,baldwin2021high,gorman2018engineering}. 

The MS and LS gates have also been applied to the multi-species atomic ions \cite{home2013quantum,tan2015multi,ballance2015hybrid,inlek2017multispecies,home2018repeated,bruzewicz2019dual,hughes2020benchmarking,wang2022significant}. The Oxford group has achieved gate fidelity of 99.8 $\%$ between a $^{43}\text{Ca}^+$ hyperfine qubit and a $^{88}\text{Sr}^+$ Zeeman qubit~\cite{hughes2020benchmarking}. %\ddd{More references for mult-species.} 

\subsection{Laser less gate}
%\ddd{Tu Hengchao} 

%Half page. Three categories. References 

%Microwave or RF field is always used to implement high-fidelity and short-time single qubit gates due to its convenient control of amplitude and phase. For multi-qubit gates, the field is required to vary significantly within the range of the ion's motion $\Delta x \sim  10 n m$ to couple the qubit and motion. However, the gradient of the microwave or RF field is proportional to 1/$\lambda \sim 1/100~ m m$, which means the Lamb-Dicke parameter $\eta \sim 10^{-7}$ is so small that the spin-motion coupling cannot be driven. A method to address this limitation is using spatially-varying-magnetic field such as static magnetic-field gradient and oscillating magnetic-field gradient.

Microwave or RF field is used to implement high-fidelity single-qubit gates with convenient control capability of amplitude and phase. In order to couple qubit and motion for multi-qubit gates, the strength of the field should vary significantly in the range of effective size of the ion, which is $z_0 \sim  10 ~ n m$.  However, the gradient of the microwave or RF field is proportional to 1/$\lambda \sim 1/100~ m m$, which is too small to drive the spin-motion interaction. In other words, the Lamb-Dicke parameter of the microwave field is around $\eta \sim 10^{-7}$, which produces negligible spin-motion coupling. A method to address this limitation is adding static magnetic fields with large gradients or using near-fields of the oscillating magnetic field, where the gradient is not limited by the wavelength of the field.

\begin{figure}[ht]
    \centering
    \includegraphics[width=0.90\linewidth]{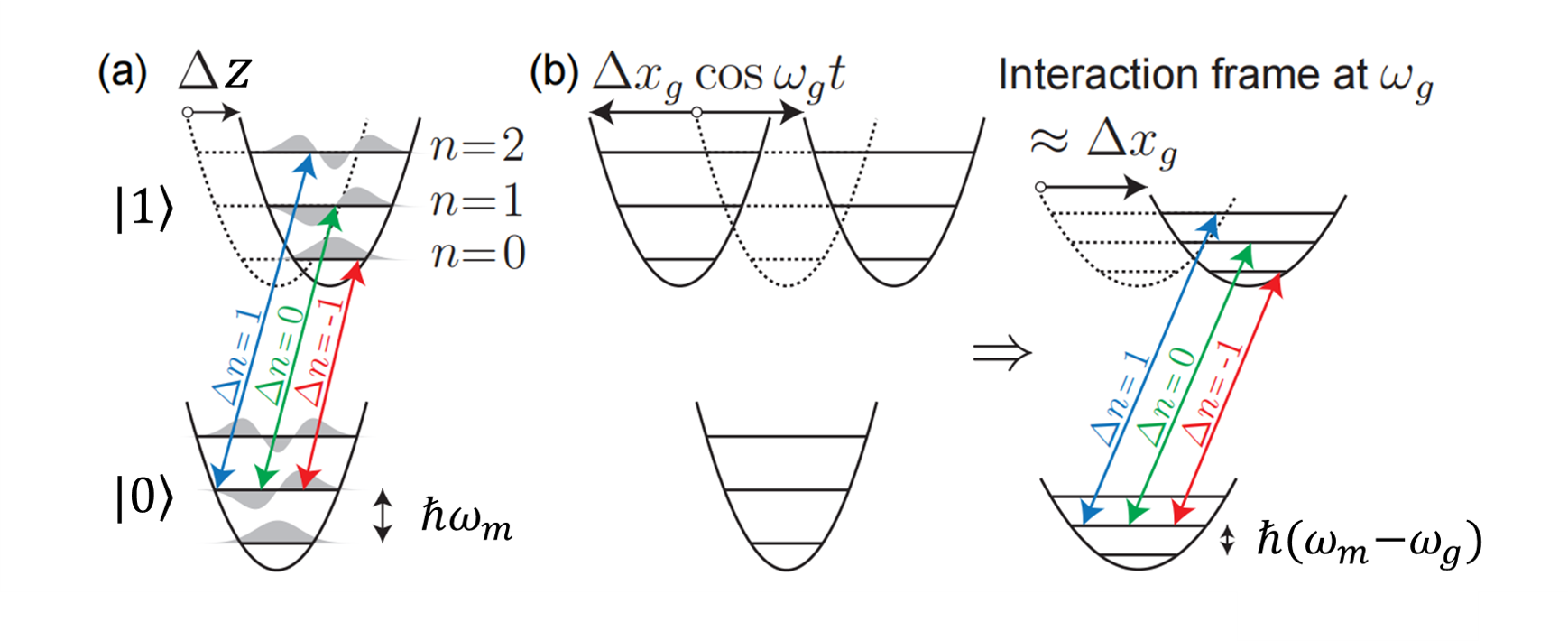}
    \caption{ Schematic description of a qubit coupled to a harmonic oscillator with a spin-dependent displacement from (a) a static magnetic field gradient or (b) an oscillating near-motion-frequency magnetic field gradient. Adapted from Ref. ~\cite{srinivas2019trapped}}.
    \label{fig:laser_less_gate}
\end{figure}

The static magnetic-field gradient used to realize spin-motion coupling was first proposed by Mintert and Wunderlich in 2001~\cite{mintert2001ion}. When an ion interacts with a static magnetic-field gradient $\frac{\partial B_z}{\partial z}$ and a near-qubit-frequency microwave, the Hamiltonian can be expressed as below,
\begin{align}
    \hat{H}     & =  \frac{1}{2}\hbar\omega_0\hat{\sigma}^z + \hbar\omega_m\hat{a}^\dagger\hat{a} \nonumber %\label{equ:LaserLessHam1} 
    \\ & + \frac{1}{2} \mu_z(B_0+\frac{\partial B_z}{\partial z} z)\hat{\sigma}^z + \hbar\Omega_{\mu}\hat{\sigma}^{x}\cos (\omega_{\mu} t ),  \label{equ:LaserLessHam2} 
\end{align}
where $\omega_0$ is the qubit frequency, $\omega_m$ is the vibrational mode frequency, $\mu_z$ is the magnetic dipole moment, $B_0$ is the magnetic field at the equilibrium position of the ion, $\frac{\partial B_z}{\partial z}$ is the gradient of the magnetic field, $z$ is the displacement around the equilibrium position, $\Omega_{\mu}$ is the rabi frequency which quantifies the strength of dipole interaction, and $\omega_u$ is the microwave near-qubit-frequency. The displacement $z$ can be changed to $z = z_{0}(\hat{a} + \hat{a}^\dagger)$, where $z_{0}=\sqrt{\frac{\hbar}{2m\omega_m }}$, is the size of ground motional state. When applying the transform $\hat{H}' = e^{\hat{S}}\hat{H}e^{-\hat{S}}$, where $\hat{S} = \eta'(\hat{a}^\dagger - \hat{a})\hat{\sigma}_z $, $\eta' = \frac{\mu_{z}z_0 }{2\hbar\omega_m} \frac{\partial B_z}{\partial z} $, the interaction part of Eq.\ref{equ:LaserLessHam2} can be expressed as~\cite{mintert2001ion,wolk2017quantum}:
\begin{align}
    \hat{H}^{'}_I     & =  \frac{1}{2}\hbar \Omega_\mu(\hat{\sigma}^{+}e^{\eta'(\hat{a}^\dagger - \hat{a})} + \hat{\sigma}^{-}e^{-\eta'(\hat{a}^\dagger - \hat{a})})(e^{i\omega_\mu t} + e^{-i\omega_\mu t}).  \label{equ:LaserLessHam3} 
\end{align}
As a consequence, the ion feels an effective Lamb-Dicke parameter $\eta'$, which means the qubit-motion coupling can be amplified by enlarging the gradient~\cite{johanning2009individual,lake2015generation}. From another perspective, if the gradient term is removed, the interaction term in Eq.~\ref{equ:LaserLessHam2} can only drive spin-flip with $\delta n = 0$. However, when introducing the gradient, the overlap between the motion states of $\ket{0}$ and $\ket{1}$ changed, which means the spin-flip with $\delta n \neq 0$ can be achieved, as shown in Fig.~\ref{fig:laser_less_gate}(a). %This phenomenon is similaer to the Franck-Condon physics~\cite{hu2011franck}. 
The static magnetic-field gradient in experiments can be produced by permanent magnets~\cite{khromova2012designer,lake2015generation,weidt2016trapped} or wires with dc current~\cite{chiaverini2008laserless,lekitsch2017blueprint,welzel2018spin}. Because of the spatially varied magnetic field, the resonance frequencies of ions are position-dependent, which provides a means to realize individual addressing by applying different microwave frequencies~\cite{johanning2009individual,wang2009individual}. 
The cross-talk of individual addressing by frequency selection was measured to be as low as $10^{-5}$~\cite{piltz2014trapped}, where the cross-talk was quantified by the number of excitations in all non-addressing qubits. The individual addressing methods were extended to the entangling gate operations, and nearest or non-nearest neighbor interactions of ions were demonstrated ~\cite{khromova2012designer}. 
%A shortcoming of the static magnetic-field gradient scheme is the necessity of using a magnetic-field sensitive qubit, where gate fidelity and coherence time can be seriously degraded by magnetic field fluctuation. A dressed state qubit with microwave, an effective clock qubit, was proposed and realized to show the coherence time is increased by two orders of magnitude to that of the bare magnetic-sensitive qubit ~\cite{timoney2011quantum}. The addressed qubits have been further developed and Then Webster demonstrated a simpler method to manipulate such dress state qubit~\cite{webster2013simple}, and Cohen introduced a high-fidelity gate proposal, which used dress state qubit in phase gate and MS gate\cite{cohen2015multi}. After that, Weidt and Randall et al. achieved ground-state with phonon number of $\overline{n} = 0.13(4)$ after sideband cooling\cite{weidt2015ground} by using dress state qubit, and they also introduced individual addressing by static magnetic field gradient and showed the efficiency of dress state qubit's preparation and detection\cite{randall2015efficient}. Finally, they achieved a two-qubit gate with high fidelity 0.985(12) and gate time 2.7$ms$\cite{weidt2016trapped}.
A shortcoming of the static magnetic-field gradient scheme is the necessity of using a magnetic-field sensitive qubit, where gate fidelity and coherence time can be seriously degraded by magnetic field fluctuation. A dressed state qubit with microwave, an effective clock qubit, was proposed and realized to show the coherence time is increased by two orders of magnitude to that of the bare magnetic-sensitive qubit ~\cite{timoney2011quantum}. The dressed qubits have been further developed and a two-qubit gate with the fidelity of  98.5~$\%$ and duration of 2.7 ms has been demonstrated \cite{webster2013simple,cohen2015multi,weidt2015ground,randall2015efficient,weidt2016trapped}.  %developed. Then Webster demonstrated a simpler method to manipulate such dress state qubit~\cite{webster2013simple}, and Cohen introduced a high-fidelity gate proposal, which used dress state qubit in phase gate and MS gate\cite{cohen2015multi}. After that, Weidt and Randall et al. achieved ground-state with phonon number of $\overline{n} = 0.13(4)$ after sideband cooling\cite{weidt2015ground} by using dress state qubit, and they also introduced individual addressing by static magnetic field gradient and showed the efficiency of dress state qubit's preparation and detection\cite{randall2015efficient}. Finally, they achieved a two-qubit gate with high fidelity 0.985(12) and gate time 2.7$ms$\cite{weidt2016trapped}.

Oscillating near-qubit-frequency magnetic field gradient was proposed as an alternative method to couple qubit and motion \cite{ospelkaus2008trapped}. Taking the gradient in the x-axis as an example, the interaction Hamiltonian can be written as:
\begin{align}
    \hat{H}_I  =\mu_x\frac{\partial B}{\partial x}x_0(\hat{a} + \hat{a}^\dagger)\hat{\sigma}_x\cos(w_{g} t),  \label{equ:LaserLessHam4} 
\end{align} 
 where $\omega_g$ is the frequency near the qubit frequency $\omega_0$. When using interaction picture with $\hat{H}_0 = \frac{1}{2}\hbar\omega_0\hat{\sigma}^z + \hbar\omega_m\hat{a}^\dagger\hat{a}$, $\hat{H}_I$ can be rewritten as:
 \begin{align}
    \hat{H}_{I}^{'}  = \hbar\hat{\sigma}_x\Omega_g \hat{a} e^{-i(\delta+\omega_m)t} + h.c.,  \label{equ:LaserLessHam5} 
\end{align} 
where $\delta = \omega_g - \omega_0$, and $\Omega_g = \frac{\mu_{x} x_0}{2\hbar}\frac{\partial B}{\partial x}$. %\ddd{make the consistent notation for operators and fidelities ($\%$).} 
Therefore, the qubit-motion coupling can be achieved. In the experiment, microwave-based two-qubit $\hat{\sigma}_{\phi}\hat{\sigma}_{\phi}$ entangling gate was performed with the fidelity of 76(3)~$\%$ ~\cite{ospelkaus2011microwave} in 2011, whose near-qubit-frequency microwave gradient was created by the current in electrodes integrated into a microfabricated trap. Individual addressing was realized with spin-flip cross-talk errors on the order of $10^{-3}$\cite{warring2013individual}. However, the fidelity of the two-qubit gate is greatly affected by residual microwave field and magnetic field fluctuation~\cite{warring2013techniques}. To alleviate these two limitations, the dynamical-decoupling method was applied to improve the fidelity to 99.7~$\%$ with a gate time of 3.25 ms \cite{harty2016high}. Another method to reduce the error from motional-mode frequency fluctuation by modulating the amplitude of the microwave was proved and realized to reach the mean fidelity of 99.7~$\%$ with the gate time of 2.938 ms \cite{zarantonello2019robust}. The method of oscillating-magnetic field gradient was integrated into a surface trap and got the two-qubit fidelity of 98.2~$\%$ ~\cite{hahn2019integrated}.

%However, it is difficult to generate the gigahertz-frequency magnetic-field gradient for a hyperfine qubit. Recently, inspired by using running optical lattice to control the ion-motion~\cite{ding2014microwave}, Srinivas and Sutherland et al. put forward a new method to couple the spin and motion, which combine near-motion-frequency oscillating magnetic field gradient with near-qubit-frequency microwave, and performed the sideband cooling to its ground state with phonon number of $\overline{n} = 0.09(7)$\cite{srinivas2019trapped}. Then they suggested a general theory with hyperfine qubits for laser less entangling gate, including above static or oscilating magnetic field gradient\cite{sutherland2019versatile}. After that, they improved the gate more robust to qubit frequency  fluctuations and motional decoherence\cite{sutherland2020laser} by intrinsic dynamical decoupling, and achieved symmetric entangled state with fidelity $1^{+0}_{-0.0017}$ and antisymmetric $0.9977^{+0.0010}_{-0.0013}$\cite{srinivas2021high}. It's the highest record in magnetic field platform so far.

However, it is difficult to generate the gigahertz-frequency magnetic-field gradient for a hyperfine qubit. Recently, inspired by using running optical lattice to control the ion-motion~\cite{ding2014microwave}, a new method to couple the spin and motion by combining near-motion-frequency oscillating magnetic field gradient with near-qubit-frequency microwave was proposed and applied to perform sideband cooling to its ground state~\cite{srinivas2019trapped}. In the scheme, the interaction Hamiltonian can be shown below:
\begin{align}
    \hat{H}_I  = \hbar\Omega_g(\hat{a} + \hat{a}^\dagger)\hat{\sigma}_z\cos(\omega_{g} t) + \hbar\Omega_{\mu}\hat{\sigma}^{x}\cos (\omega_{\mu} t ),  \label{equ:LaserLessHam6} 
\end{align} 
where $\Omega_g = \frac{\mu_{z} x_0}{2\hbar}\frac{\partial B_z}{\partial x}$, $\omega_g$ is the frequency near the motion frequency $\omega_m$, $\omega_\mu$ is the frequency near the qubit frequency $\omega_0$, and $\Omega_{\mu}$ is the Rabi frequency of magnetic dipole interaction. Transforming to the interaction picture with respect to $\hat{H}_0 = \hbar\omega_g\hat{a}^\dagger\hat{a}$ and drop the fast term, the total Hamiltonian will change to:
\begin{align}
    \hat{H}    &= \frac{1}{2}\hbar\omega_0\hat{\sigma}^z + \hbar(\omega_m-\omega_g)\hat{a}^{\dagger}\hat{a} \label{equ:LaserLessHam7} 
    \\ & + \frac{1}{2}\hbar\Omega_g\hat{\sigma}^z(\hat{a} + \hat{a}^{\dagger}) + \hbar\Omega_{\mu}\hat{\sigma}^{x}\cos (\omega_{\mu} t ),  \label{equ:LaserLessHam8} 
\end{align} 
which is similar to the Hamiltonian in static magnetic field gradient method, except changing the $\omega_m$ to $\omega_m - \omega_g$ in Eq.~(\ref{equ:LaserLessHam2}) showed in Fig.~\ref{fig:laser_less_gate}(b). In the experiment, a robust gate to qubit frequency  fluctuations and motional decoherence was proposed by using intrinsic dynamical decoupling~\cite{sutherland2020laser}.
The experimental realization of the gate produced a symmetric entangled state with near-perfect fidelity and antisymmetric with $99.77~\%$~\cite{srinivas2021high}.

%Then they suggested a general theory with hyperfine qubits for laser less entangling gate, including above static or oscilating magnetic field gradient\cite{sutherland2019versatile}.

%- Using magnetic field gradient
%\cite{timoney2011quantum,khromova2012designer,bermudez2012robust,belmechri2013microwave,tan2013demonstration,webster2013simple,mikelsons2015universal,cohen2015multi,weidt2016trapped,lekitsch2017blueprint}

%- Microwave gradient
%\cite{ospelkaus2011microwave,warring2013individual,warring2013techniques,carsjens2014surface,sutherland2019versatile}

%- Dzmitry way - combinations
%\cite{ding2014microwave,sutherland2019versatile,srinivas2019trapped,srinivas2021high}

\section{IV. Discussion about the control methods}
%\ddd{Cai Zhengyang\ Ou Lingfeng}
%\ppp{ We may need to discuss why we need the modulation? We need to talk about constraints and how to fulfill the constraints? It is not yet written.}

\begin{figure*}[tb]
\center
\includegraphics[width=18cm]{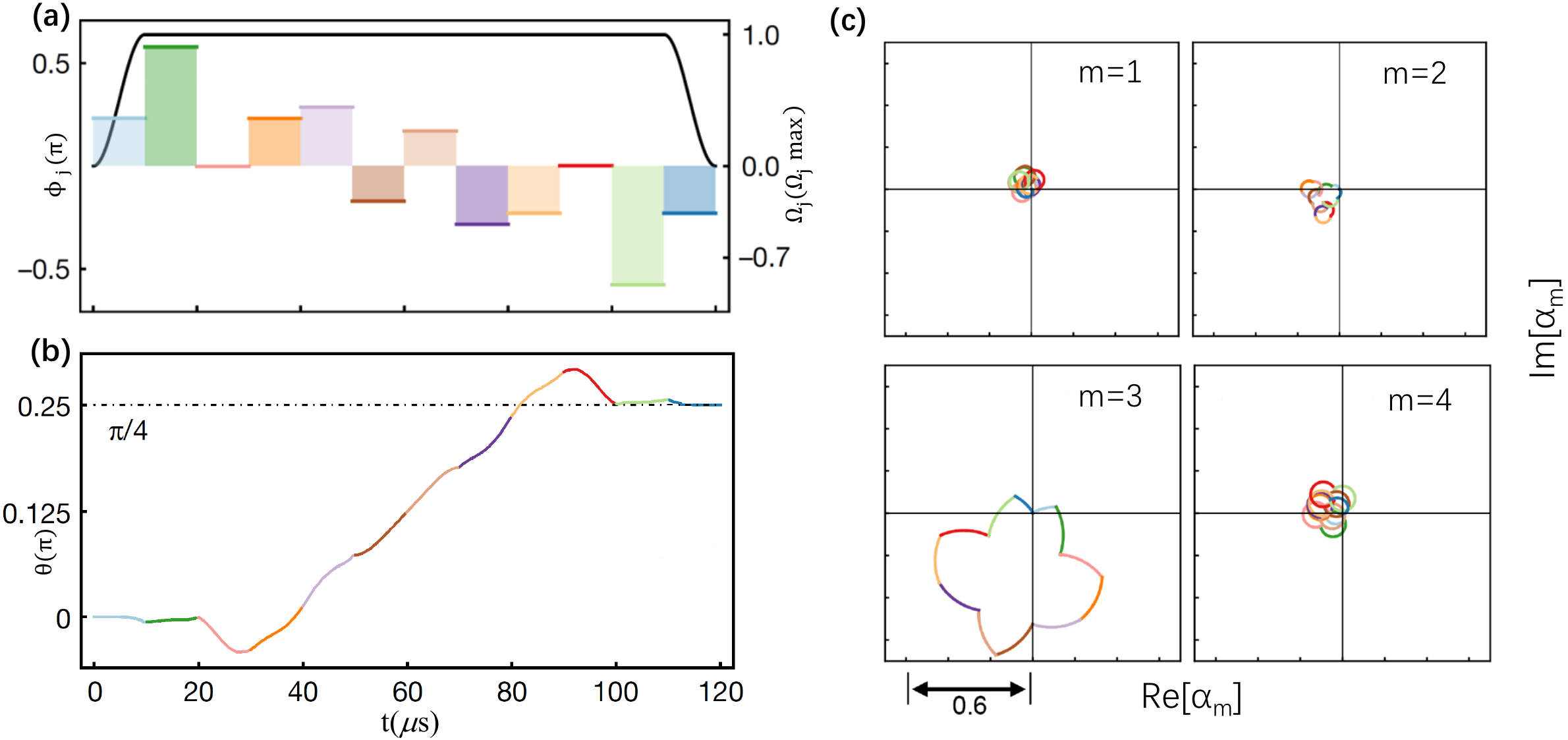}
\caption{Scheme for an entangling operation on the two ion-qubits in the middle of a four-ion chain. (a) The modulation scheme of the pulse sequence in motional phase $\phi_j$ and the amplitude, the Rabi frequency $\Omega_j$, which are corresponding to the colored segments and the black solid curve respectively. (b) Accumulation in the geometric phase $\theta$ on the middle two ion-qubits. (c) Phase-space trajectories of four vibration modes.  Modified from$~$\cite{lu2019global}}
\label{FIGURE_IV.1}
\end{figure*}

%In pursuit of faster gate speed, in quantum gate operation, the internal state of the ion is coupled with multiple vibrational modes through spin-dependent force.%

Recently many modulation schemes have been developed for the entangling gate operations with trapped ions. The main purposes to include modulations in the gate can be categorized as the following three : (1) to make the gate robust against experimental imperfections, (2) to speed up the gate, and (3) to implement the gates in a single trap with multiple ions.     

%After the gate operation is completed, a perfect maximum entanglement gate satisfies two constraints: 1. The curve enclosed by the phase space corresponding to each vibration mode of the system is closed $|\alpha_j(\tau_g)|=i\frac{\eta_{j,m}\Omega}{2}\int_0^\tau e^{i\delta_m t}dt= 0$(FIG. \ref{FIGURE_IV.1}). 2. The sum of the areas enclosed by the phase space curves of the system in different modes is $\theta={\pi }/{4}+2m\pi $, where $m$ is an integer. However, due to the experimental errors, for example, the frequency fluctuation of driving laser, the bight of the phonon mode can not lock as designed. They decrease the fidelity of gates. To increase the robustness of gates against errors, people modified the operation laser. Besides, the pursuit of short gate time and scaling the gate in multi-qubit systems are also the motivation for applying the modulation-laser scheme. The laser-control methods include amplitude \cite{roos2008ion,benhelm2008towards,kirchmair2009deterministic,zarantonello2019robust,zhu2006arbitrary,zhu2006trapped,choi2014optimal,steane2014pulsed,schafer2018fast}, phase \cite{green2015phase,milne2020phase}, frequency \cite{leung2018robust,wang2020high, kang2021batch}, multi-tone \cite{haddadfarshi2016high,webb2018resilient, shapira2018robust, shapira2020theory}, Walsh modulation \cite{hayes2012coherent,srinivas2021high}, and laser dynamics decoupling \cite{manovitz2017fast}.  

%robust gate
As discussed in the previous section, a perfect two-qubit entangling gate should satisfy two constraints at the end of the gate: (a) The trajectory in the phase space of the vibrational mode of an ion chain is closed. (b) The areas enclosed by the phase-space trajectory in the mode are $\theta={\pi }/{4}$. However, due to the experimental imperfections such as frequency fluctuations of the driving field, timing error of the gate, and amplitude fluctuation, the trajectory of the vibrational mode can not be closed as designed and the accumulated geometric phase may not be exactly $\pi/4$, which decreases the fidelity of gates. The robustness of gates against imperfections can be improved by including modulations on the driving fields. 

%As discussed in the previous section, a perfect two-qubit entangling gate should satisfy two constraints at the end of the gate: (a) The trajectory in the phase space of the vibrational mode of an ion chain is closed as shown in FIG. \ref{FIGURE_IV.1}.c. (b) The sum of the areas enclosed by the phase-space trajectories in different modes is $\theta={\pi }/{4}$, which is displayed as FIG. \ref{FIGURE_IV.1}.b. However, due to the experimental imperfections such as frequency fluctuations of driving field, the bight of the phonon mode can not be closed as designed, which decreases the fidelity of gates. The robustness of gates against imperfections can be improved by including modulations on the control fields. 

The duration of two-qubit gates can be shortened by increasing the amplitude of Rabi-frequency $\Omega$ in Eq.~(\ref{MS Hamiltonian}). If the strength of sideband transitions $\eta\Omega$ is comparable to the frequency difference of two vibrational modes, which are the center of mass (COM) and the stretch modes, we cannot use the single mode description discussed in the previous section. It is necessary to include the effects of both vibrational modes for the two-qubit gate operations. That is, the phase-space trajectories of both modes should be  closed and the sum of the areas enclosed by the phase-space trajectories in both modes should be ${\pi}/{4}$ \cite{garcia2003speed,duan2004scaling,steane2014pulsed,taylor2017study,torrontegui2020ultra,wang2022fast}. These requirements can be achieved by using the modulation methods and experimentally realized~\cite{wong2017demonstration,schafer2018fast}.

One promising scheme to scale up the number of ions for quantum computation is to perform quantum gates on ion qubits confined in a long ion chain \cite{zhu2006arbitrary,zhukas2021high,lin2013sympathetic}. In the single-trap scheme, the transverse modes are more popularly used than axial modes. It is mainly due to the fact that better laser cooling can be achieved in transverse modes since the frequencies of transverse modes are larger than those of axial modes in the linear chain with a large number of ions~\cite{lin2009large,kim2009entanglement}. However, the frequency spacings between the modes are getting smaller as the number of ions increases in the single linear trap, which makes it challenging to address only a single vibrational mode for the entangling gate. Similar to the fast gate, it is necessary to close all the trajectories of the modes and make the sum of the phase to be $\pi/4$ for target ions as shown in Fig.~\ref{FIGURE_IV.1}.  

%\ddd{multiple ions in a single}
%and scaling the gate in multi-qubit systems are also the motivation for applying the modulation-laser scheme. 

%The laser-control methods include amplitude \cite{roos2008ion,benhelm2008towards,kirchmair2009deterministic,zarantonello2019robust,zhu2006arbitrary,zhu2006trapped,choi2014optimal,steane2014pulsed,schafer2018fast}, phase \cite{green2015phase,milne2020phase}, frequency \cite{leung2018robust,wang2020high, kang2021batch}, multi-tone \cite{haddadfarshi2016high,webb2018resilient, shapira2018robust, shapira2020theory}, Walsh modulation \cite{hayes2012coherent,srinivas2021high}, and laser dynamics decoupling \cite{manovitz2017fast}. 

In order to achieve these goals, many different types of modulation such as amplitude \cite{roos2008ion,benhelm2008towards,kirchmair2009deterministic,zarantonello2019robust,zhu2006arbitrary,zhu2006trapped,choi2014optimal,steane2014pulsed,schafer2018fast,lin2009large}, phase \cite{hayes2012coherent,green2015phase,manovitz2017fast,milne2020phase,bentley2020numeric,srinivas2021high}, frequency modulation \cite{leung2018robust,leung2018robust,leung2018entangling,landsman2019two,wang2020high,kang2021batch}, and multi-frequency have been developed \cite{haddadfarshi2016high,webb2018resilient, shapira2018robust, shapira2020theory}. Here we briefly overview the developments of those modulations and discuss the multi-frequency modulation can provide a general method covering all the other modulation methods.  

%The ideal geometry-phase gate is independent of the temperature\GGG{(why need to mention temperature )}, but it is sensitive to the parameters of the laser (or the microwave) calibration. In the experiment, we will encounter various problems to reduce the fidelity of the geometry-phase gate, the dephasing caused by the heating of the motion mode, the frequency detuning, the gate time error, and the influence of the carrier part. To improve gate fidelity and robustness to noise, modulation schemes are needed, which include amplitude\cite{roos2008ion,benhelm2008towards,kirchmair2009deterministic,zarantonello2019robust,zhu2006arbitrary,zhu2006trapped,choi2014optimal,steane2014pulsed,schafer2018fast}, phase\cite{green2015phase,milne2020phase}, frequency\cite{}, multi-tone\cite{}, Walsh modulation\cite{hayes2012coherent,srinivas2021high}, and laser dynamics decoupling\cite{manovitz2017fast}.%

%With unmodulated lasers (or microwaves) for gate operation, there is no guarantee that both conditions are met simultaneously, which results in a loss of fidelity. However, one can use modulated lasers (or microwaves) to satisfy both constraints. This ensures high fidelity when chasing a fast speed. %

\subsection{Amplitude}
%\ddd{Ou Lingfeng}
%Half page test2.

The amplitude modulation method is a scenario that  changes the amplitude of the driving field, $\Omega$ in Eq.~(\ref{MS Hamiltonian}) from a constant to a time-dependent sequence during the gate operation.

\begin{figure}[ht]
    \centering    \includegraphics[width=1\linewidth]{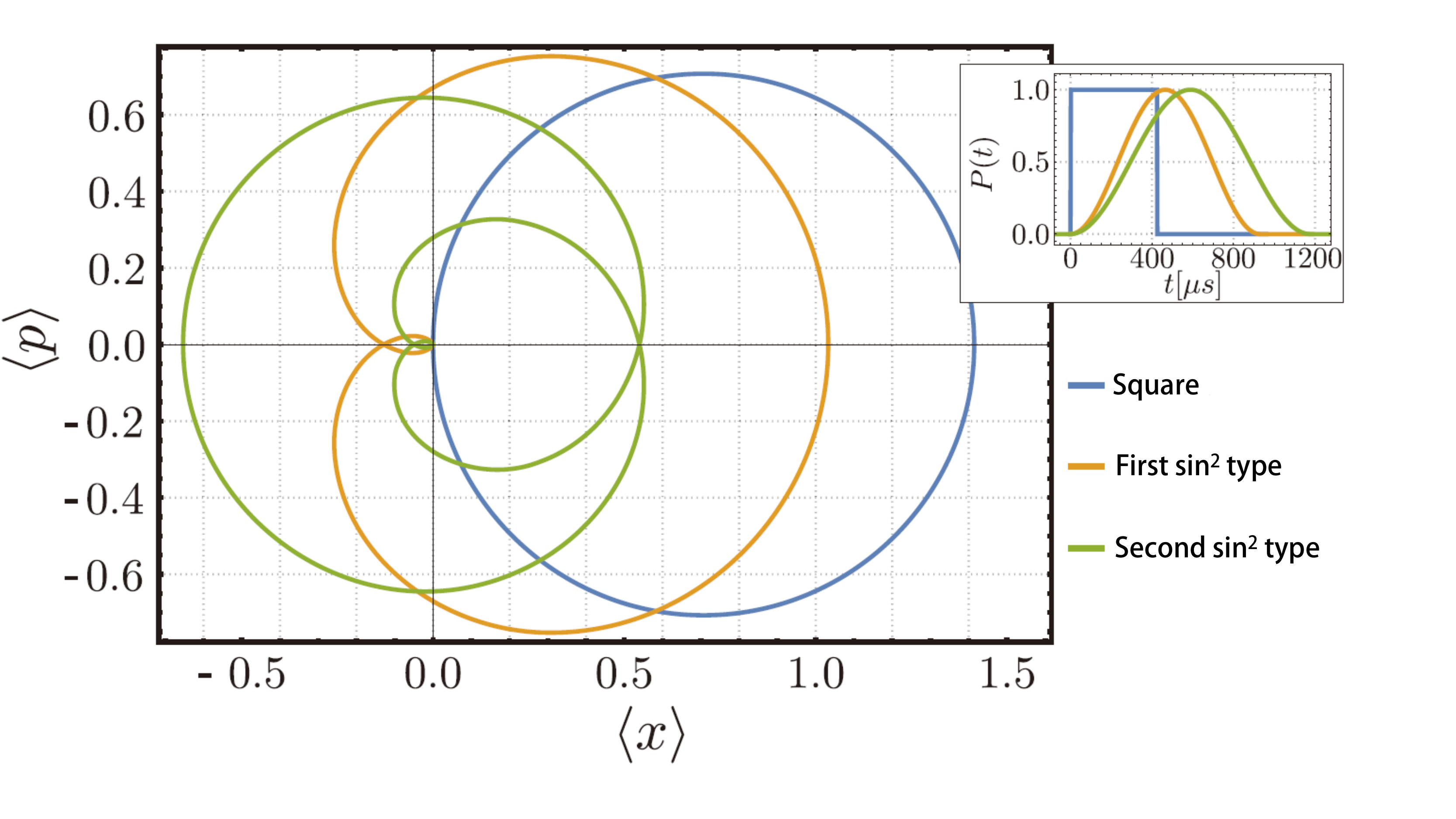}
    \caption{The phase-space trajectories in the case of three amplitude-modulation schemes, which are square pulse (blue), first (orange) and second (green) ${{\sin }^{2}}$ type sequence shown in the inset \cite{zarantonello2019robust}. }
    \label{FIGURE_IV.2}
\end{figure}

Continuous-amplitude modulation scheme for two-qubit gates was proposed for robustness to the optical phase fluctuation of driving beams~\cite{roos2008ion}. This scheme was applied to MS gates on optical qubits of $^{40}\rm{Ca}^{+}$ ions with the duration of $50\mu s$ and the fidelity of $99.3\%$~\cite{benhelm2008towards}. The amplitude-shaped pulses were also used to realize MS gates on the ions in the thermal state with average phonon number $\bar{n} = 20$, which showed the fidelity of $97.4~\%$ in $25~\mu s$ duration~\cite{kirchmair2009deterministic}. The continuous amplitude modulation is also applied to the MS gates by using a near-field microwave~\cite{zarantonello2019robust}. It was confirmed that this scheme is robust to frequency fluctuations of the vibrational modes. The gate was implemented with hyperfine qubits in $^{9}\rm{Be}^{+}$ ions with $99.7~\%$ fidelity \cite{zarantonello2019robust}. The pulse timing and consequent phase space trajectory are as shown in Fig.~\ref{FIGURE_IV.2}. %During their research, they all used only one phonon mode as the bus during the quantum gates operation.

The discrete amplitude modulation was proposed to realize two-qubit gates with multiple vibrational modes~\cite{zhu2006arbitrary,zhu2006trapped,lin2009large}. In the scheme, the laser field was divided into isochronous segments with different intensities to fulfill the constraints discussed above for ideal gates. This approach was implemented using hyperfine $^{171}\rm{Yb}^{+}$-ion qubits in a five-ion chain with gate time and fidelity of $190\mu s$ and $95\%$, respectively~ \cite{choi2014optimal}. Another scheme of discrete amplitude modulation scheme was proposed to regard the duration and amplitude of each optical pulse as variables~\cite{steane2014pulsed}. The optimal pulse sequence is considered to be insensitive to the optical phase of driving laser~\cite{steane2014pulsed}. The above scenario was implemented in the hyperfine qubits of $^{43}\rm{Ca}^{+}$ ions. The implemented two-bit gate has $99.8~\%$ fidelity during $1.6~\mu s$~\cite{schafer2018fast}. Additionally, there are applications of the amplitude modulation gates, such as programmable quantum computer~\cite{debnath2016demonstration}, and parallel entangling gates~\cite{figgatt2019parallel,grzesiak2020efficient}.

\subsection{Phase}
%\ddd{Ou Lingfeng}
%\ddd{what would be the good transitions for the phase modulations}

The phase modulation (PM) method is to modulate the vibrational phases $\phi_j$ in Eq.~(\ref{MS Hamiltonian}) of the spin-dependent force during the gate operation. The PM has an advantage in experimental realization since it can be more precisely controlled in the experiment than amplitude. The first PM was proposed to alternating the vibrational phases between 0 and $\pi$ in the form of the Walsh functions for suppressing effects of a certain frequency and timing errors, which was realized with $^{171}{\rm Yb}^{+}$ ~\cite{hayes2012coherent}. This scheme does not require optimizing the pulse sequence, but the duration of the gate increases as the higher orders of Walsh functions are used for enhanced robustness. The Walsh function method was also utilized in mixed-species \cite{hughes2020benchmarking} and  $^{40}\rm{Ca}^{+}$ \cite{clark2021high} ions system. Similar to the Walsh function method, a phase modulation scheme based on the dynamical decoupling pulses technique has been proposed and implemented, which improved the robustness against dephasing noise of the qubit without much time overhead of the dynamical decoupling pulses~\cite{manovitz2017fast}.

A more flexible piecewise-constant PM scheme with continuous values of phases was proposed to implement two-qubit gates for both robustness and high fidelity with multiple vibrational modes ~\cite{green2015phase}. This scheme was implemented in the hyperfine qubit of two $^{171}\rm{Yb}^{+}$ ions, which demonstrated the robustness of this scheme to static or time-varying errors of laser amplitude and detuning. Their two-qubit achieves an average $99.4\%$ fidelity in about $310~\mu s$ \cite{milne2020phase}.

We note that compared with amplitude modulation, it can change the orientation of the trajectory in phase space promptly, which is more proper to prevent the vibrational state of ions from being excited beyond the Lamb-Dicke regime and to implement a fast gate~\cite{wang2022fast}.

%\ddd{Advantages of PM : precise implementation is possible.} 
%\ddd{Walsh modulation as a kind of phase modulation}

%\ddd{Kaizhao discussion about phase modulation}

\subsection{Frequency}
%\ddd{Cai Zhengyang} 
%Half page

%\ddd{better definition for FM} %By changing the instantaneous frequency of the field, frequency modulation (FM) can encode information about the carrier. 
%\ddd{Advantagnes and disadvantages}
%In contrast to FM, discrete PM methods require exponentially more pulses as the number of ions increases if the individual motion modes are to be decoupled~\cite{green2015phase}. 
The frequency modulation (FM) method is to modulate the laser frequency $\omega$, which can be considered as $\mu$ above the Eq.~(\ref{MS Hamiltonian}). The FM was introduced for the robustness of the two-qubit gates with multiple vibrational modes~\cite{leung2018robust}. The FM can be conceptually equivalent to the PM, but in the experimental realizations, wider ranges of frequencies are modulated, which cannot be simply converted to the PM. The FM is implemented by directly adjusting the frequency of the driving field through Acusto-Opic Modulator (AOM), but it will also change the deflection angle of the beam, causing errors in the offset of the beam focus point.

\begin{figure}[ht]
    \centering
    \includegraphics[width=0.90\linewidth]{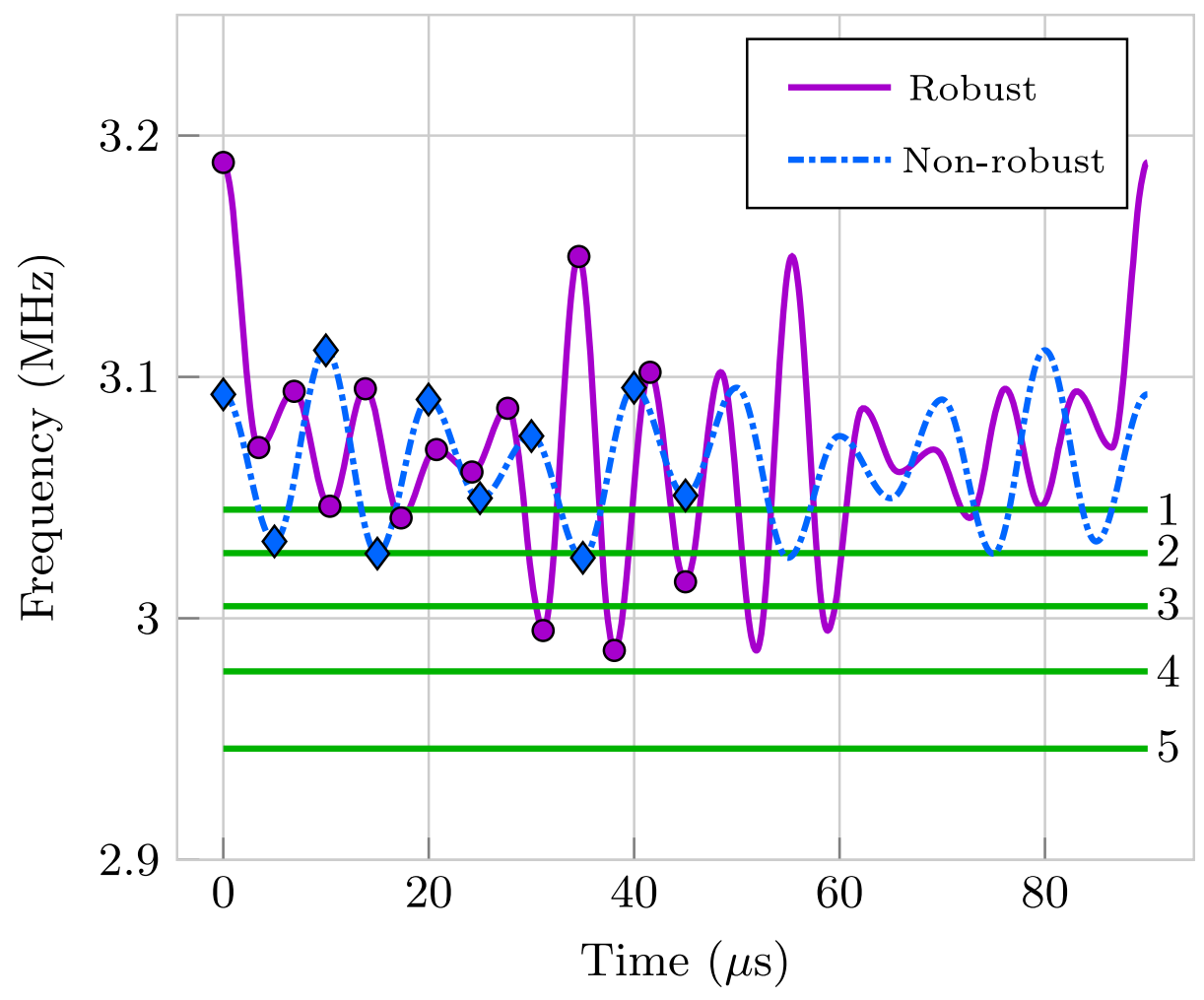}
    \caption{Frequency modulation pulse for 2-qubit gate in 5-ion chain, the green lines are the mode frequencies.   \cite{leung2018robust}. }
    \label{FIGURE_7}
\end{figure}
%A continuous frequency-modulated laser was developed to minimize the time average of the amplitude of the trajectory function $|\alpha_k^j(\tau)|\propto|\int_0^\tau\int_0^t e^{-i\theta_k(t')}dt|$ , making it robust to frequency error and achieving 98.3(4)\% fidelity in experiments\cite{leung2018robust}.

A continuous frequency-modulated laser was developed to minimize the residual spin-motion entanglement at the end of the two-qubit gate, making it robust to frequency error and achieving 98.3 \% fidelity in experiments with 5 ions~\cite{leung2018robust} as shown in Fig.~\ref{FIGURE_7}. %\ddd{is there a comparison to amplitude and phase modulation, continuous phase modulation, easier to implement than discrete modulation? It will be difficult as the number of ions increases. Any disadvantage? The technical problem of position shifts...May also put Q-Ctrl paper?}

With two individually addressed beams focused on the ion chain, discrete frequency modulation was applied to realize the two-qubit gate fidelity of 99.49 \% in the 2-ion chain and 99.30 \% fidelity in the 4-ion chain in 200 $\mu s$~\cite{wang2020high}. Inspired by machine learning, the frequency modulation gate was optimized by using a large sample set and mini-batches. The batch-optimized frequency-modulation gate reached the gate fidelity of 99.08 \% in 120 $\mu s$~\cite{kang2021batch}.

Apart from the work with a single modulation, there are also progresses with two modulations, such as amplitude modulation plus frequency modulation~\cite{leung2018entangling, landsman2019two}, amplitude modulation plus phase modulation~\cite{lu2019global, bentley2020numeric,srinivas2021high}.
%The Walsh modulation has been applited to mitigate the impact of symmetry frequency errors of driving laser at the expense of increasing gate time, which was realized in the two-$^{25}\rm{Mg}^{+}$-ions system with $1_{-0.0017}^{+0}$ fidelity in $740\mu s$ \cite{srinivas2021high}.

Table \ref{Table.1} summarizes some representative works of recent experimental realization in two-qubit gates.
\begin{table*}
	\begin{center}
		\centering
		\caption{Summary of experimental realizations for two-qubit gates. }
		\begin{tabular*}{\textwidth}{@{\extracolsep{\fill}}lcccccccc}
		   \hline\rule{0pt}{10pt}Modulation&Ions&Qubit &Control method &Ion number &Fidelity &Gate time &Year&Reference\\
	       \hline\rule{0pt}{11pt}Amplitude&$^{40}\rm{Ca}^{+}$&Optical&Laser&Two&$99.3\%$&$50\mu s$ &2008&\cite{benhelm2008towards}\\
	      
	       Amplitude&$^{171}\rm{Yb}^{+}$&Hyperfine&Laser&Five &$95.0\%$&$190\mu s$&2014&\cite{choi2014optimal}\\
	       Amplitude&$^{43}\rm{Ca}^{+}$&Hyperfine&Laser&Two &$99.8\%$&$1.6\mu s$&2018&\cite{schafer2018fast}\\
	       Amplitude&$^{9}\rm{Be}^{+}$&Hyperfine&Microwave&Two&$99.7\%$&$\sim 3000\mu s$&2019&\cite{zarantonello2019robust}\\
	       
	       Phase&$^{43}\rm{Ca}^+, ^{88}\rm{Sr}^+$&Hyperfine, Zeeman&Laser&Two&$99.8\%$&$49.2\mu s$&2020&\cite{hughes2020benchmarking} \\
	       Phase&$^{171}\rm{Yb}^{+}$&Hyperfine&Laser&Two &$99.4\%$&$\sim 310\mu s$&2020&\cite{milne2020phase}\\
	       Phase&$^{25}\rm{Mg}^{+}$&Hyperfine&Microwave&Two&$1_{-0.0017}^{+0}$&$740\mu s$&2021&\cite{srinivas2021high}\\
	       Phase&$^{40}\rm{Ca}^+$&Optical&Laser&Two&$99.94\%$&$35\mu s$&2021&\cite{clark2021highfidelity} \\
	       
	       Frequency&$^{171}\rm{Yb}^{+}$&Hyperfine&Laser&Five &$98.6\%$&$90\mu s$&2018&\cite{leung2018robust}\\
           Frequency&$^{171}\rm{Yb}^{+}$&Hyperfine&Laser&Four &$99.30\%$&$200\mu s$&2020&\cite{wang2020high}\\

           None&$^{171}\rm{Yb}^{+}$&Dress state&Microwave&Two &$98.5\%$&$2.7m s$&2016&\cite{weidt2016trapped}\\
           % None&$^{25}\rm{Mg}^{+}$&Hyperfine&Near-qubit-frequency RF gradient&Two &$76\%$&Noon&2011&\cite{ospelkaus2011microwave}\\
           None&$^{43}\rm{Ca}^{+}$&Hyperfine&Microwave&Two &$99.7\%$&$3.25m s$&2016&\cite{harty2016high}\\
           None&$^{43}\rm{Ca}^+$&Hyperfine&Laser&Two&$99.9\%$&$\sim 100\mu s$&2016&\cite{ballance2016high} \\
           None&$^{9}\rm{Be}^+$&Hyperfine&Laser&Two&$99.92\%$&$\sim 30\mu s$&2016&\cite{gaebler2016high} \\
           None&$^{9}\rm{Be}^+$&Hyperfine&Microwave&Two&$97.4\%$&$\sim 105\mu s$&2018&\cite{tan2018demonstration} \\
           None&$^{9}\rm{Be}^{+}$&Hyperfine&Microwave&Two &$98.2\%$&$808\mu s$&2019&\cite{hahn2019integrated}\\
           None&$^{171}\rm{Yb}^+$&Hyperfine&Laser&Two&$99.8\%$&$30\mu s$&2020&\cite{baldwin2020subspace} \\
			\hline
		\end{tabular*}
		\label{Table.1}
	\end{center}
\end{table*}

\subsection{Multi-frequency Methods} 
%\ddd{Cai Zhengyang} 
%Half page
%- Brief description of multi-frequency method, pointing out it covers amplitude, phase, and frequency modulations. 

The multi-frequency modulation method realizes the gate by simultaneously driving multiple-motional modes using the laser field of multiple frequencies with different amplitudes. It can provide the general and systematic control methods including all the effects of amplitude, phase, and frequency modulation.

%A multi-frequency method is a scheme to realize gates by driving ions with a multi-chromatic laser field. It can achieve all the effects of amplitude, phase, and frequency modulation, and can also have more (?) degrees of freedom to achieve more constraints. We take the MS gate as a example.

% \begin{equation}
%     H_{MS}=(\gamma(t)a+\gamma^*(t)a^\dagger)S_x
% \end{equation}

%\ddd{KK }
% %the MS gate as a example.
% Where $S_x=\sum_j \sigma_x^{(j)},\gamma(t)=\eta\Omega e^{i\delta t}$. If we add modulation, then the $\gamma_m=\eta \Omega e^{i\delta t}$, which can be expanded into a Fourier series.
% \begin{align}
%     \gamma_m    & = \eta\Omega(t)e^{i\int_0^t\delta(t')dt'}\\\
%                 & = \sum_{j=1}^m c_j e^{ij\omega t}
% \end{align}

The field for MS gates on $j$-th ion can be simplified as $ \gamma_j(t) = \Omega_j e^{-i\mu t}e^{i \phi_{\rm m, \it j}}$ as shown in Eq.~(\ref{MS Hamiltonian}). The amplitude, the phase, and the frequency of the field are $\Omega_j $, $\phi_{\rm m,\it j}$, and $\mu$, respectively. We can generate an arbitrary waveform of the field with amplitude, phase, and frequency modulation by letting them the functions of time, which can be decomposed using Fourier expansion as
\begin{equation}
    \gamma_j(t)=\sum_{n=-\infty}^{\infty} \Omega_{n,j } \exp(-i n \omega t), 
\end{equation}
where $\Omega_{n,j}$ is a complex amplitude of $n^{\rm th}$ component of the frequency $n \omega$, and $\omega=\frac{2\pi}{\tau_g}$ with $\tau_g$ at the duration of the gate. This Fourier expansion can be seen as applying multi-frequency components. From the completeness of the Fourier series, we find the generality of the multi-frequency method as an arbitrary modulation with amplitude, phase, and frequency.%, which was proposed in Ref. ~\cite{haddadfarshi2016high,leung2018robust,webb2018resilient,shapira2020theory}. 

% \begin{equation}
%     H_{MS}=\hbar\delta S_x\sum_{j=1}^N c_j(a^\dagger e^{ij\delta t}+a e^{-ij\delta t})
% \end{equation}

% In 2016, haddadfarshi et.al proposed a multi-frequency gate and gave an implementation of a global entangling gate $e^{-i\frac{\pi}{8}S_x^2}$, and it pointed out that the phase trajectory of a multi-tone gate is closer to closure under non-ideal conditions~\cite{haddadfarshi2016high}.

The multi-frequency method was proposed to improve the gate performance and implement a global entangling gate~\cite{haddadfarshi2016high}. It pointed out that the phase trajectories of a multi-tone gate are closer to closure under non-ideal conditions than single-tone gate~\cite{haddadfarshi2016high}. These multi-frequency MS gates were realized with \Yb~and \Sr~ions in experiments and the robustness of the multi-frequency gates was experimentally verified against gate time error and frequency detuning drift~\cite{webb2018resilient,shapira2018robust}. The multi-frequency method was further developed to include general constraints for robustness conditions to mitigate the effects of frequency drift, gate time offset, and carrier coupling to achieve a robust global entangling gate~\cite{shapira2020theory,manovitz2022trapped}. In the next section, we will discuss the details of the multi-frequency methods for robustness, speed-up, and multi-qubit operations. 

\section{V. Multi-frequency Method}
%\ddd{Zhang Jing-Ning}
%\ddd{KK brief summary of this section for the introduction}

In this section, first, we review the trapped-ion gates based on various state-dependant forces in detail, mainly for hyperfine qubits, and discuss the general framework of the multi-frequency methods. Finally, we consider the application of multi-frequency methods for fast and global entangling gates with and without individual addressing capability of the systems. 

\subsection{Various types of spin-dependent force}
\begin{figure}
\centering
\includegraphics[width=0.95\linewidth]{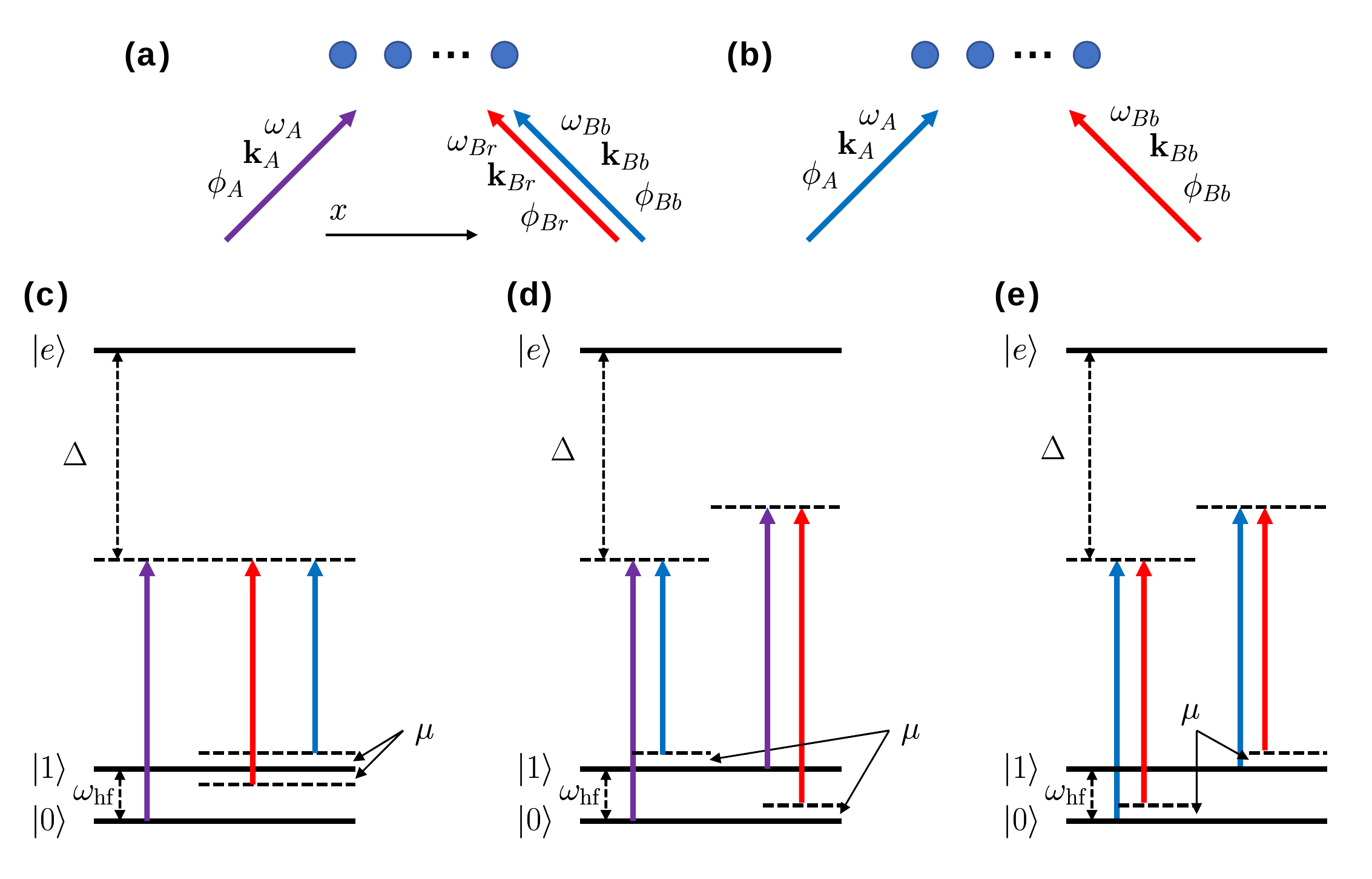}
\caption{Laser geometries. (a) Two non-copropagating Raman lasers shining on a trapped-ion chain.  One of the two Raman lasers is monochromatic while the other is bichromatic. The detuning is around the qubit-state transition frequency. The direction of the net transferred wave vector is assumed to be along the axial direction. (b) The same as (a) with two monochromatic Raman laser beams. The detuning is around the frequencies of relevant motional modes. (c, d) Relevant energy levels and laser frequencies for the phase-sensitive gate (c) and the phase-insensitive gate. (e) The same as (c, d) for the $\hat\sigma^z$ gate. }
\label{fig:laser geometries}
\end{figure}

We first briefly recall the theory of the entangling gates based on the spin-dependent forces, covering both cases of the $\hat\sigma_\phi$ gate and the $\hat\sigma_z$ gate~\cite{lee2005phase}. As mentioned above, transitions between qubit states can be driven by non-copropagating Raman lasers. Here we consider Raman couplings mediated by an excited electronic state $\ket{e}$, with laser configurations shown in Figs.~\ref{fig:laser geometries} (a, b), and assume the net wave vector of the Raman process is parallel to the $x$-axis of the trap. To mitigate the spontaneous emission, the detuning $\Delta$ from the excited state is large, so the condition $\Delta\gg\omega_{\rm hf}$ holds good in both cases. The interaction Hamiltonian between the ions and multiple lasers is written as follows (set Planck constant $\hbar = 1 $),
\begin{eqnarray}
\hat H(t)&=&\sum_i\sum_{\alpha}\sum_{s\in\left\{0,1\right\}}g_{\alpha,i,s}\cos\left({\mathbf k}_\alpha\cdot {\mathbf r}_i-\omega_\alpha t-\phi_{\alpha,i}\right)\\
&&\times\left(\ket{e}\bra{s}_i+\ket{s}\bra{e}_i\right),\nonumber
\end{eqnarray}
with ${\mathbf k}_\alpha$ and $\omega_\alpha$ being the wave vector and the frequency of the corresponding Raman laser. With the capability of single-ion addressing, it is possible to tune independently the single-photon Rabi frequencie $g_{\alpha,i,s}$ and the phase $\phi_{\alpha,i}$, which provides enough freedom to construct multi-qubit entangling operations including the global gate \cite{leibfried2005creation,monz201114}.

For the case of the $\sigma_\phi$-gate, the qubit states $\ket{0}$ and $\ket{1}$ can be encoded with hyperfine levels insensitive to the magnetic field, i.e. the clock states, which have the same single-photon Rabi frequencies, $g_{\alpha,i,0}=g_{\alpha,i,1}\equiv g_{\alpha,i}$. Transitions between the qubit levels can be induced by a monochromatic (subscripted by "A") and a bichromatic (subscripted by "Br" and "Bb") Raman lasers, as shown in Fig.~\ref{fig:laser geometries} (a), with the frequency difference of the two Raman beams being around the hyperfine splitting $\omega_{\rm hf}$ and the detuning of the stimulated Raman process to the carrier transition denoted by $\mu$. If the frequency of Raman laser A is larger or smaller than both of the two frequencies in Raman laser B, the spin part of the effective laser-ion interaction after eliminating the excited state will depend on the optical phase difference between the Raman lasers. As a result, this type of entangling gate is called the phase-sensitive gate. (for example see Fig.~\ref{fig:laser geometries} (c), where the frequencies of the Raman lasers can be set as $\omega_{Br}=\omega_A-\omega_{\rm hf}+\mu$ and $\omega_{Bb}=\omega_A-\omega_{\rm hf}-\mu$) After eliminating the excited state and introducing the Lamb-Dicke approximation, the effective laser-ion interaction Hamiltonian in the rotating frame is
\begin{eqnarray}
\hat H_I(t)&=&\sum_m\omega_m\hat a_m^\dag\hat a_m+\cos\mu t\label{eq:phase_sensitive}\sum_i\Omega_i\label{ep:H_phase_sensitive}\\
&&\times\left[\hat\sigma_i^{\phi_i}+
\sum_m\eta_{i,m}\left(\hat a_m^\dag+\hat a_m\right) \hat\sigma_i^{(\phi_i-\frac{\pi}{2})}\right],\nonumber
\end{eqnarray}
where the effective Rabi frequency and the spin phase are $\Omega_i=\frac{g_{A,i}g_{Br,i}}{2\Delta}$ and $\phi_i=\phi_{A,i}-\phi_{Br,i}-\delta k\bar x_i$, respectively, where we assume $g_{Br,i}=g_{Bb,i}$ and $\phi_{Br,i}=\phi_{Bb,i}$. 
Here the net-transferred wave vectors $\delta k\hat{\mathbf e}_x={\mathbf k}_A-{\mathbf k}_{Br}={\mathbf k}_A-{\mathbf k}_{Bb}$ are along the $x$-direction with the unit vector $\hat{\mathbf e}_x$ and $\bar x_i$ is the equilivrium position of the $i$-th ion. The site- and mode-resolved Lamb-Dicke parameters are $\eta_{i,m}=\eta_mb_{i,m}$ with $\eta_m=\delta k\sqrt{\frac{\hbar}{2M\omega_m}}$ and $b_{i,m}$ being the element of the matrix that diagonalizes the collective motion of the ion crystal.

On the other hand, if the frequency of Raman laser $A$ is lying in between those of the Raman laser B, as shown in Fig.~\ref{fig:laser geometries} (d), the effective ion-laser interaction Hamiltonian becomes,
\begin{eqnarray}
\hat H_I(t)&=&\sum_m\omega_m\hat a_m^\dag\hat a_m+\sum_i\Omega_i\Big[\cos\left(\mu t-\phi_i\right)\label{eq:sigma_x_sdf}\label{eq:H_phase_insensitive}\\
&&-\sin\left(\mu t-\phi_i\right)\sum_m\eta_{i,m}\left(\hat a_m^\dag+\hat a_m\right)\Big]\hat\sigma_i^x.\nonumber
\end{eqnarray}
In this case, the spin part is independent of the phase difference of the Raman laser beams, thus leading to the phase-insensitive gate.

Besides the $\hat\sigma_\phi$-dependent force, it is also possible to induce $\hat\sigma_z$-dependent force using magnetically sensitive hyperfine states, with the laser configuration shown in Fig.~\ref{fig:laser geometries} (b), where there are both monochromatic lasers along the $A$ and $B$ optical paths. With the laser frequencies shown in Fig.~\ref{fig:laser geometries} (e), i.e. $\omega_B=\omega_A-\mu$, the effective Hamiltonian becomes
\begin{eqnarray}
\hat H_I(t)&=&\sum_m\omega_m\hat a_m^\dag\hat a_m+\sum_i\Omega_i\Big[\cos\left(\mu t-\phi_i\right)\nonumber\\
&&-\sin\left(\mu t-\phi_i\right)\sum_m\eta_{i,m}\left(\hat a_m^\dag+\hat a_m\right)\Big]\hat\sigma_i^z.\label{eq:H_light_shift}
\end{eqnarray}
%\begin{eqnarray}
%\hat H_I(t)&=&\sum_m\omega_m\hat a_m^\dag\hat a_m+\sum_i\Omega_i\cos\left(\delta kx_i+\mu t-\phi_i\right)\hat\sigma_i^z\nonumber\\
%&=&\sum_m\omega_m\hat a_m^\dag\hat a_m+\sum_i\Omega_i\Big[\cos\left(\mu t-\phi_i\right)\nonumber\\
%&&-\sin\left(\mu t-\phi_i\right)\sum_m\eta_{i,m}\left(\hat a_m^\dag+\hat a_m\right)\Big]\hat\sigma_i^z,\label{eq:sigma_z_sdf}
%\end{eqnarray}
%where the relation between the parameters of the spin-dependent force and those of the Raman lasers are
%\begin{eqnarray}
%&&\Omega_i=\frac{\Omega_{i,0}-\Omega_{i,1}}{2},\\
%&&\Omega_{i,0}=\frac{g_{A,i,0}g_{B,i,0}}{2\left(\Delta+\omega_{\rm hf}\right)},
%\quad\Omega_{i,1}=\frac{g_{A,i,1}g_{B,i,1}}{2\Delta},\nonumber\\
%&&\delta k{\mathbf e}_x={\mathbf k}_{A,i}-{\mathbf k}_{B,i},\quad\phi_i=\phi_A-\phi_B-\delta k\bar x_i.\nonumber
%\end{eqnarray}
Note that in this case the effective Rabi frequency $\Omega_i$ depends on the differential AC-Stark effect, and thus this scheme only works for magnetically sensitive ion qubits. There is an alternative proposal for the light-shift or $\hat\sigma_z$ gate with clock-state qubits~\cite{aolita2007high,roos2008ion,kim2008geometric}, which has been adopted by the Honeywell ion-trap group~\cite{baldwin2021high}.

%\subsection{General framework of multi-frequency method}
\subsection{General framework of multi-frequency method}

For trapped-ion systems, the periods of the trapping potential provide a natural time scale for quantum operations. The characteristic time scales for the prementioned entangling gates involving a single motional mode are much longer than the trapping period. On the contrary, fast gates are operated on a time scale comparable with the trapping period. Intuitively, the laser intensities, as well as the Rabi frequencies, are large in order to exert enough influence on the trapped-ion system in such a short operation time, which results in non-negligible off-resonant carrier coupling and invalidation of the Lamb-Dicke approximation.

To derive fast gate schemes, we focus on the phase-insensitive $\hat\sigma^x$ gate and the light-shift $\hat\sigma^z$ gate. Despite the different laser configurations, the derivation can be unified by substituting $\hat\sigma_i^x$ in Eq.~(\ref{eq:H_phase_insensitive}) and $\hat\sigma_i^z$ in Eq.~(\ref{eq:H_light_shift}) with $\hat\sigma_i^\alpha$ ($\alpha=x,$ or $z$). The reason that we do not consider the phase-sensitive gate is that the effect of the carrier coupling cannot be canceled due to the $\pi/2$ phase difference between the carrier and the qubit part of the sideband terms. It is obvious from the effective Hamiltonians in Eqs.~(\ref{eq:H_phase_insensitive}) and (\ref{eq:H_light_shift}) that the detuning $\mu$ of the stimulated Raman process introduces an intensity modulation function in the spin-dependent force. In the slow region where the gate length is much longer than the trapping period, the detuning $\mu$ can be chosen to be close to a single motional mode, such that only this mode is excited and the others can be effectively eliminated in the dynamics. In this case, the monochromatic modulation will suffice to disentangle the qubit state and the motional modes. In the fast regime, however, the frequency differences between $\mu$ and multiple motional modes are comparable with the Rabi frequency $\Omega$, so multiple motional modes take part in the dynamics and need to be disentangled. To tackle this complicated situation, Refs.~\cite{haddadfarshi2016high,shapira2020theory,wang2022fast} proposed a multichromatic modulation scheme, which, compared to the prementioned monochromatic modulation scheme, introduces more controlling parameters to satisfy multiple constraints imposed by the requirements both to disentangle the motional modes and to apply effective spin-spin interactions. 

In the Lamb-Dicke regime, the effective Hamiltonian with multichromatic modulation in the rotating frame is written as follows,
\begin{eqnarray}
\hat H_I(t)&=&\sum_{i=1}^N\Big[f_{{\rm car},i}(t)+f_{{\rm sdf},i}(t)\label{eq:effective_Hamil}\\
&&\times\sum_{m=1}^N\eta_{i, m}\left(\hat a_me^{-i\omega_mt}+\hat a_m^\dag e^{i\omega_mt}\right)\Big]\hat\sigma_i^\alpha,\nonumber
\end{eqnarray}
with $\alpha\in\left\{x,z\right\}$ and the multichromatic modulation functions for the carrier and the spin-dependent force terms being
\begin{eqnarray}
f_{{\rm car},i}(t)&=&\Omega_i\sum_{k=1}^K r_{i,k}\cos\left(\nu_kt-\phi_{i,k}\right),\label{eq:modulation_functions}\\
f_{{\rm sdf},i}(t)&=&\Omega_i\sum_{k=1}^K r_{i,k}\sin\left(\nu_kt-\phi_{i,k}\right),\nonumber
\end{eqnarray}
where $\Omega_i$ is the strength of the spin-dependent force and $\vec{r}$ is a normalized dimensionless $K$-entry vector ($\left\|\vec r_i\right\|_2=1$) characterizing the distribution of laser power among different frequency components. The frequency components are determined by the gate duration $\tau$, such that $\nu_k=2k\pi/\tau$ ($k=1,\ldots,K$), which guarantees that the contribution of the carrier coupling always vanishes at the end of the dynamics. For the derivation to be succinct and clear, we set $\phi_{i,k}=0$ for $\forall i$ and $\forall k$ irrespective of any position and frequency dependence that may exist in real systems. Later for the generality, we will consider gate schemes that are robust with respect to the overall fluctuation of the optical phases. 

Using the Magnus expansion, we obtain the evolution operator $\hat U(t)$ as follows,
\begin{eqnarray}
\hat U(t)=\exp\left[-i\sum_{i,m}\hat B_{i,m}(t)\hat\sigma_i^\alpha+i\sum_{i<j}\Theta_{i,j}(t)\hat\sigma_i^\alpha\hat\sigma_j^\alpha\right],\label{eq:unitary_individual}
\end{eqnarray}
with $\hat B_{i,m}(t)=\beta_{i,m}^*(t)\hat a_m+\beta_{i,m}\hat a_m^\dag$, where $\beta_{i,m}(t)$ and $\Theta_{i,j}$ can be obtained as
\begin{eqnarray}
\beta_{i,m}(t)&=&\eta_{i,m}\int_0^t f_{{\rm sdf},i}(t')e^{i\omega_m t'}dt',\\
\Theta_{i,j}(t)&=&2\sum_m\eta_{i,m}\eta_{j,m}\int_0^t\int_0^{t'} f_{{\rm sdf},i}(t')\nonumber\\
&&\times f_{{\rm sdf},j}(t'')\sin\omega_m\left(t'-t''\right)dt'dt''.
\end{eqnarray}
To construct a target unitary operator with effective spin-spin interactions, i.e. $\hat U_{\rm tar}(\tau)=\exp\left(-i\sum_{i<j}J_{i,j}\hat\sigma_i^\alpha\hat\sigma_j^\alpha\right)$, we obtain the following constraints to be satisfied at time $t=\tau$,
\begin{eqnarray}
\beta_{i,m}(\tau)=0,\quad \Theta_{i,j}(\tau)=J_{i,j}.
\end{eqnarray}
For an $N$-ion system, the number of constraints is $N^2+\frac{N(N-1)}{2}$. Thus solutions exist as long as the variational modulation scheme provides more controlling parameters. Moreover, as variational modulation schemes always provide much more degrees of freedom, it is possible to search for optimized schemes with respect to realistic physical considerations. For example, we can require the maximum Rabi frequency to be as small as possible by optimizing a cost function  ${\mathcal C}\left(\left\{\Omega_i,\vec{r}_i,\vec\phi_i\right\}\right)={\rm max}_i\left|\Omega_i\right|$.

%\subsection{Application to Multi qubits gate without individual control}
\subsection{Framework of multi-frequency method with global control}

In general trapped ion systems, individual control requires highly focused laser beams with the beam waist smaller than the spatial separation of ions. Thus it not only introduces complicated experimental instruments to generate and control the individual laser beams but also introduces extra errors in the manipulation procedure of the quantum system. For example, tiny movements in the ion position will lead to serious amplitude fluctuation in the Raman process, due to the steep profile of the highly-focused beams. Moreover, the residual electromagnetic field experienced by the neighboring ions also induces coherent crosstalk errors. Thus the entangling-gate schemes that ease the requirement for individual control always attract considerable attention.

Without individual control, the control parameters are reduced as $\Omega_i\equiv\Omega$, ${\vec r}_i\equiv\vec r=\left(r_1,\ldots,r_K\right)$ and ${\vec \phi}_i\equiv\vec\phi=\left(\phi_1,\ldots,\phi_K\right)$, with $K$ being the number of frequency components. We further assume $\phi_k=\phi$ for simplicity. In this symmetric case, it is natural to define the collective spin operators $\hat S_m^\alpha=\sum_{i=1}^Nb_{i,m}\hat\sigma_i^\alpha$ and the evolution operator in Eq.~(\ref{eq:unitary_individual}) is reduced to the following form,
\begin{eqnarray}
\hat U(t)&=&\exp\Big[-i\sum_{m=1}^N\left(\hat a_mA_m^*(t)+{\rm h.c.}\right)\hat S_m^\alpha\nonumber\\
&&+i\sum_{m=1}^N\Theta_m(t)\hat S_m^{\alpha2}\Big],\label{eq:evol_op}
\end{eqnarray}
with
\begin{eqnarray}
A_m(t)&=&\eta_m\Omega\int_0^t f_{\rm sdf}(t')e^{i\omega_mt'}dt',\\
\Theta_m(t)&=&\eta_m^2\Omega^2\int_0^t\int_0^{t'}f_{\rm sdf}(t')f_{\rm sdf}(t'')\nonumber\\
&&\times\sin\left[\omega_m\left(t'-t''\right)\right]dt'dt''.
\end{eqnarray}
For the all-to-all entangling gate, we required that at the end of the dynamics $t=\tau$ the following constraints are satisfied,
\begin{eqnarray}
&&A_m\left(\tau\right)=0,\quad\forall m,\\
&&\Theta_1(\tau)=\pi/4,\quad \Theta_{m>1}=0,\nonumber
\end{eqnarray}
where the first part is the aforementioned motion-spin disentangling condition. For target unitary that represents a general spin-spin interaction, $\hat U(\tau)=\exp\left(-\sum_{i<j}J_{i,j}\hat\sigma_i^\alpha\hat\sigma_j^\alpha\right)$,
the effective interaction strength $J_{i,j}$ can be obtained as
\begin{eqnarray}
J_{i,j}=2\sum_m\Theta_m(\tau)b_{i,m}b_{j,m}.\label{eq:linear_equations}
\end{eqnarray}
In other words, for a given set of $J_{i,j}$, gate schemes without individual control exist only when the linear-equation system in Eq.~(\ref{eq:linear_equations}) has non-vanishing solutions.

\subsection{Considerations for fast and robust global-entangling gate}
When the gate length is comparable to the trapping period, the above gate scheme becomes sensitive to the fluctuation of the optical phases and the drift of the equilibrium ion positions. These effects result in an indefinite initial phase in the modulation function. Moreover, the required Rabi frequencies, which are proportional to the laser intensities, also increase as the gate length becomes shorter. Thus some of the motional modes are strongly driven during the evolution, and the Lamb-Dicke approximation may not always hold good. In other words, the higher-order terms of the Lamb-Dicke parameters cannot be neglected. In this case, a protocol obtained within the Lamb-Dicke approximation may suffer beyond-Lamb-Dicke error in experiment realization. Here we introduce two sets of constraints to guarantee the robustness of the resulting gate schemes with respect to the initial phase and higher-order terms.

\begin{figure}
\centering
\includegraphics[width=0.95\linewidth]{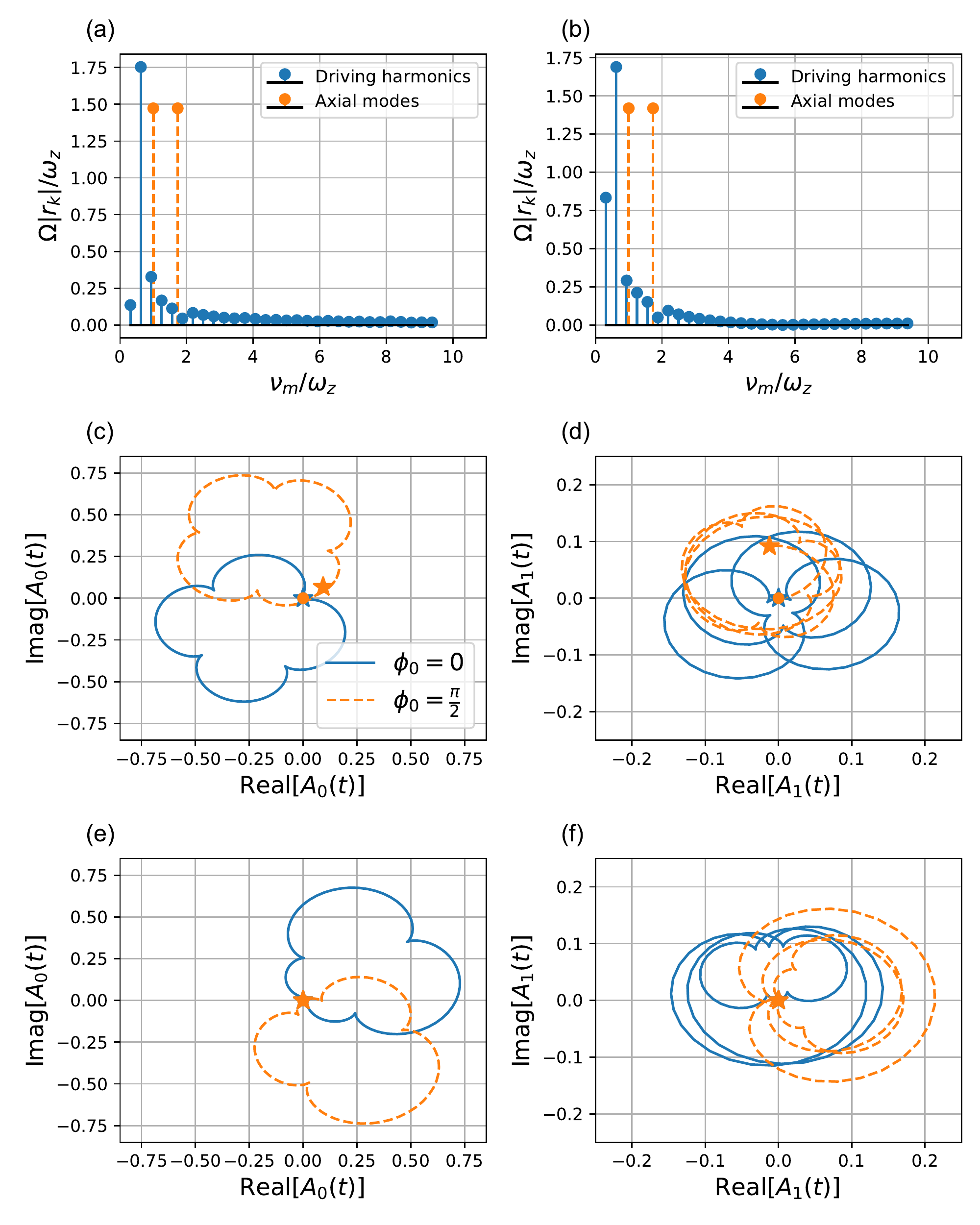}
\caption{Two-qubit entangling gate robust with respect to the initial phase drift. Here we assume the lasers are coupled to the axial-$z$ modes, with the center-of-mass mode frequency $\omega_0$ equal to the axial trapping frequency $\omega_z=2\pi\times 1$ MHz. The gate duration is $\tau=3.2~\mu$s and $K=30$ different driving harmonics with the frequency $\nu_k=2k\pi/\tau$, $k=1,2,\ldots K$ are considered. (a) Driving amplitudes for the gate scheme without considering the robustness against the initial phase drift.  The phase-space trajectories of the center-of-mass mode (c) and the stretch mode (d) are shown for different values of the initial phase $\phi_0$. (b, e, f) The same as (a, c, d) for a robust gate scheme considering the initial phase drift. In (c-f), the endpoints of the trajectories at the end of the gate operations are marked with stars. The gate schemes are obtained under the condition $\phi_0=0$, and only for the robust gate, the endpoints return to the origin for a drifted initial phase $\phi_0=\pi/2$.}
\label{fig:robust_gate}
\end{figure}

To obtain robust gate schemes with respect to the drift of the initial phase, we notice that the modulation function in Eq.~(\ref{eq:modulation_functions}) can be decomposed as $f_{\rm sdf}(t)\equiv f_s(t)\cos\phi+f_c(t)\sin\phi$, with
\begin{eqnarray}
f_s(t)=\Omega\sum_{k=1}^Kr_k\sin\nu_kt,\quad f_c(t)=\Omega\sum_{k=1}^Kr_k\cos\nu_kt.
\end{eqnarray}
Note that before we obtain gate schemes for $\phi=0$, where we consider linear constraints with $f_{\rm sdf}(t)=f_s(t)$. Requiring the linear constraints $A_m(\tau)=0$ be satisfied for arbitrary $\phi$ is equivalent to adding another set of linear constraints with $f_{\rm sdf}(t)=f_c(t)$. Similarly, we decompose the quadratic constraints for $\Theta_m(\tau)$ into four sets of constraints. Figure~\ref{fig:robust_gate} shows two-qubit gate schemes in a two-ion chain with the axial trapping frequency $\omega_z = 2\pi\times 1$ MHz. Without loss of generality, we assume the axial modes are driven and the gate duration is set to be $3.2~\mu$s. As expected, the robust gate scheme obtained in the above procedure is insensitive to the drift of the initial phase, as the endpoints of the trajectories of both motional modes come back to the origin at the end of the gate operation, as shown in Fig.~\ref{fig:robust_gate} (e, f). Under the Lamb-Dicke approximation, the fidelity of the gate scheme without consideration of the initial phase drift decreases to $98.9\%$ with $\phi_0 = \pi/2$, while the fidelity of the robust gate is always perfect irrespective of the initial phase. The cost of the robustness is, however, that the dimensionless magnitude of the Rabi frequency $\left|\Omega\right|/\omega_z$ increases from about 1.81 to 1.93. The amount of the increment becomes severe for gate schemes with shorter gate durations.

To suppress the beyond-Lamb-Dicke error, it is a natural way to first expand the effective Hamiltonian in Eq.~(\ref{eq:effective_Hamil}) to the second order of the Lamb-Dicke parameters $\eta_m$. In this case, the evolution operator in Eq.~(\ref{eq:evol_op}) contains second-order terms, which depend linearly on the modulation function according to the Magnus expansion. As a result, there will be another set of linear constraints to be satisfied. Although errors induced by the second-order terms are suppressed, the resulting gate schemes always have much higher Rabi frequencies than the original gate scheme.  As a result, these schemes only work well in the ultrafast regime, where the beyond-Lamb-Dicke errors are significant. 

%\ddd{include the example of global gate with a figure}

\subsection{Comments on the individual control capability for multi-qubit gates}

On the contrary, with the ability to individually address each qubit, the multi-qubit entangling gate scheme can be more flexible. For example, the global entangling gate scheme via individual phase modulation in Ref.~\cite{lu2019global} can be reduced to a similar gate involving fewer ion qubits by simply keeping the individual lasers shining on the selected ions and turning off the others. Intuitively, gate schemes with individual control are usually more efficient than those without individual control for  entangling gates only involving a small part of the ions in a multi-qubit system. However, for a concrete target gate, the compromise between individual and global control needs to be considered seriously. 

\section{VI. Conclusion and Outlook}
%\ddd{K Kim}

In this article, we review the quantum gates for trapped-ion quantum computation and quantum simulation. For the single qubit gates, it has been shown that the duration of the gates approach to picoseconds and the fidelities much higher than typical error correction requirements have been demonstrated with both microwaves and laser beams. For the two-qubit entangling gates, the main speed limit comes from the frequency of the vibrational modes, and the gates close to the limit have been demonstrated. The fidelities of the gates are limited by photon scatterings, heating and dephasing of the vibrational modes, dephasing of qubits, amplitude fluctuations of the laser fields, and so on. In principle, these errors can be further suppressed to below the level of $10^{-4}$ infidelity. Currently, one big challenge is to realize the gates with such high fidelity in a scalable way, either using ion shuttling or multiple ions in a single trap. One interesting quantum gate with trapped ions is the global gates that contain equivalently $\approx N^2/2$ gates in a single operation \cite{leibfried2005creation,monz201114,lu2019global}. If a global gate can be realized at a speed close to the trap frequency, then the speed per two-qubit gates can be considered as the same factor $\approx N^2/2$ of improvement.

%Related to errors. 
%Qudit?

$Note$- while preparing the manuscript, we learned that Ref.~\cite{yum2022progress} reviewed the trapped-ion quantum gates, similar contents to this paper. 

%\textit{Note added.--}

\begin{acknowledgments}
This work was supported by the innovation Program for Quantum Science and Technology under Grants No. 2021ZD0301602, and the National Natural Science Foundation of China under Grants No.92065205, and No.11974200.
\end{acknowledgments}

%\bibliography{refs}

%

\end{document}